\documentclass{article}

\usepackage{PRIMEarxiv}

\usepackage[utf8]{inputenc} 
\usepackage[T1]{fontenc}    
\usepackage{hyperref}       
\usepackage{url}            
\usepackage{booktabs}       
\usepackage{amsfonts}       
\usepackage{nicefrac}       
\usepackage{microtype}      
\usepackage{lipsum}
\usepackage{graphicx}
\usepackage[caption=true,labelformat=simple]{subfig}
\usepackage{siunitx}
\usepackage{lineno}
\usepackage{booktabs}
\usepackage{flafter} 
\usepackage{tcolorbox}
\usepackage{xspace}
\usepackage{adjustbox}
\usepackage{standalone}
\usepackage{tikz}
\usepackage{tabularx}
\usepackage{placeins}
\usepackage{amssymb}
\usepackage{amsmath}
\graphicspath{{Images/}}     

\newcommand{\gf}{\texttt{Geant4}\xspace}
\newcommand{\iminuit}{\texttt{iminuit}\xspace}

\title{ Simulation Studies of the Effect of  SiPM Dark Noise on the Performance of a Highly Granular Crystal ECAL}

\author{
  J.Rolph\thanks{{corresponding author}} , Y.\,Liu, B.\,Qi\\ \textbf{and on behalf of the CEPC Working Group} \\
  Institute of High Energy Physics \\
  Chinese Academy of Sciences \\
  Yuquan Road 19B \\
  Beijing 100049 \\ 
  China\\
  \texttt{jackrolph@ihep.ac.cn} \\
   \And
  Z. Zhao  \\
  Shanghai Jiao Tong University \\
  800 Dongchuan Road \\ 
  Shanghai 200240 \\ 
  China \\
}

\begin{document}
\maketitle

\begin{abstract}
A proposal for the CEPC ECAL is a highly-granular scintillating crystal design that uses SiPMs to measure physics signals from photons. Radiation damage to the silicon will impair the performance of the calorimeter due to dark noise, which will affect the reconstruction capabilities of the calorimeter system. This paper presents a simulation study assessing the effect of radiation damage of SiPM dark noise on the response from calorimeter to electrons due to changing fluence and temperature. It was observed that dark noise significantly degrades the llnearity of response, with up to \qty{45}{\percent} error in reconstructed energy for a \qty{1}{\giga \electronvolt} shower at a fluence of \qty{1e10}{\per \centi \meter \squared}. The stochastic and noise resolution terms was observed to remain stable, increasing only by \qty{0.2}{\percent} and \qty{1}{\percent} respectively in the range \qtyrange{1e7}{1e10}{\per \centi \meter \squared} fluence. Under the assumption of no irradiation, the influence of dark noise with temperature in the normal operating range of the calorimeter system was estimated to be negligible.
\end{abstract}


\section{Introduction}

Future electron-positron collider experiments such as the Circular Electron Positron Collider (CEPC), Future Circular Collider-electron (FCC-ee), International Linear Collider (ILC), and Compact Linear Collider (CLIC), will precisely measure the properties of the Higgs boson, weak vector bosons, and the top quark, while also performing critical validations of the Standard Model. A typical figure of merit for the required performance is a Particle Flow is to reconstruct the mass of the Higgs boson with \qtyrange{3}{4}{\percent} resolution. This approach necessitates accurate tracking of charged particles within a jet, sophisticated event reconstruction techniques, and highly granular sampling calorimeters. Furthermore, the electromagnetic calorimeter (ECAL) must possess a resolution sufficient to reconstruct the energy of photons, while also being designed with enough granularity to resolve individual particles with excellent energy and time resolution.

One proposal for the design of the CEPC ECAL system is a high-granularity scintillating crystal design, which can achieve energy resolutions better than \qty{3}{\percent} while maintaining a highly granular readout and minimising dead volume. In this design, the photons produced by the crystal are detected by silicon photomultipliers (SiPMs). SiPMs are photodetectors that operate in the near-infrared to ultraviolet frequency range. They can count single photons, exhibit high photon-detection efficiency compared to standard photomultipliers, and can operate in magnetic fields. These characteristics make SiPMs a frequently chosen readout sensor for highly granular detector designs.


The endcap regions of the calorimeter system in future electron-positron collider experiments will be subjected to extreme radiation conditions. High-energy protons, photons, pions, and neutrons with kinetic energies exceeding \qty{1}{\kilo \electronvolt} will undergo non-ionising energy loss (NIEL) as they traverse silicon, which can displace one or more atoms in the crystal lattice. These ‘defects’ can introduce additional energy levels in the silicon bandgap. Defects facilitate the thermal excitation of charge carriers, leading to the generation of leakage current via Shockley-Reid-Hall generation-recombination \cite{garutti_radiation_2019, acerbi_understanding_2019}. The subsequent leakage current can induce random, uncorrelated Geiger discharges in the SiPM called 'dark noise.' The Poole-Frenkel effect \cite{frenkel_pre-breakdown_1938} and thermally enhanced trap-assisted tunnelling through the silicon bandgap \cite{vincent_electric_1979} also contribute to dark noise.

Dark noise will increase with fluence as more defects are introduced over time, adversely affecting the calorimeter's performance to an unknown extent. For CEPC, simulation studies on fluence are not yet available. This study aims to estimate the influence of radiation damage on the resolution and noise trigger efficiency of the crystal ECAL exposed to varying levels of fluence, utilising simulations of dark noise from a SiPM and a typical physics signal from the calorimeter. Additionally, an estimation of the relationship between temperature fluctuations and dark count rate is studied. Additionally, the influence of the temperature on the detector resolution, assuming no irradiation, is estimated. 


In the text, $\Phi$ is fluence in units of \unit{\per \centi \meter \squared}. DCR and $DCR$ are the dark count rates as concepts and a variable, respectively. The variable is measured in \unit{\hertz} or \unit{\mega \hertz}. Where appropriate, the $DCR$ is sometimes presented normalised to the area of an SiPM of area \qtyproduct{3 x 3}{\milli \meter \squared}, in units of \unit{\hertz \per \milli \meter \squared} or \unit{\mega \hertz \per \milli \meter \squared}.

\section{Methods and Tools}

\subsection{Highly-Granular Crystal ECAL Prototype}

Each layer of the crystal ECAL consists of 40 scintillating crystal bars per layer, each of \qtyproduct{40 x 1 x 1 }{\centi \meter} volume. Each subsequent layer is rotated orthogonally with respect to the previous one. This 'cross-hatch' structure enables reconstruction of the spatial position of energy deposits in the crystal by combining information from two orthogonal and adjacent layers and transverse granularity while minimising dead volume and the number of readout channels compared to a design with single crystals. Scintillation light from the crystal is read out from each end by two SiPMs.  In this study, the crystal is assumed to be beryllium germanium oxide (BGO). The layers are arranged in 'supercells' such that the calorimeter has lateral dimensions of \qtyproduct{40 x 40}{\centi\meter} and longitudinal dimensions of \qty{28}{\mathrm{layers}} or around \qty{24}{\ensuremath{X_{0}}}, where $\mathrm{X}_{0}$ is the radiation length of the crystal. Each end of each bar has a SiPM readout, meaning there is a total of \numproduct{28 x 40 x 2} (2240) SiPMs per 'standard' module, or 1120 SiPMs for each side of every bar. The properties of the SiPM that are considered are shown in Appendix Table \ref{tab:SiPMSimulationParameters}. It is noted that this number is not fixed, as the geometry must vary depending on where the module is located in the final detector. 

\begin{figure}[htpb]
    \centering
    \includegraphics[width=0.8\linewidth]{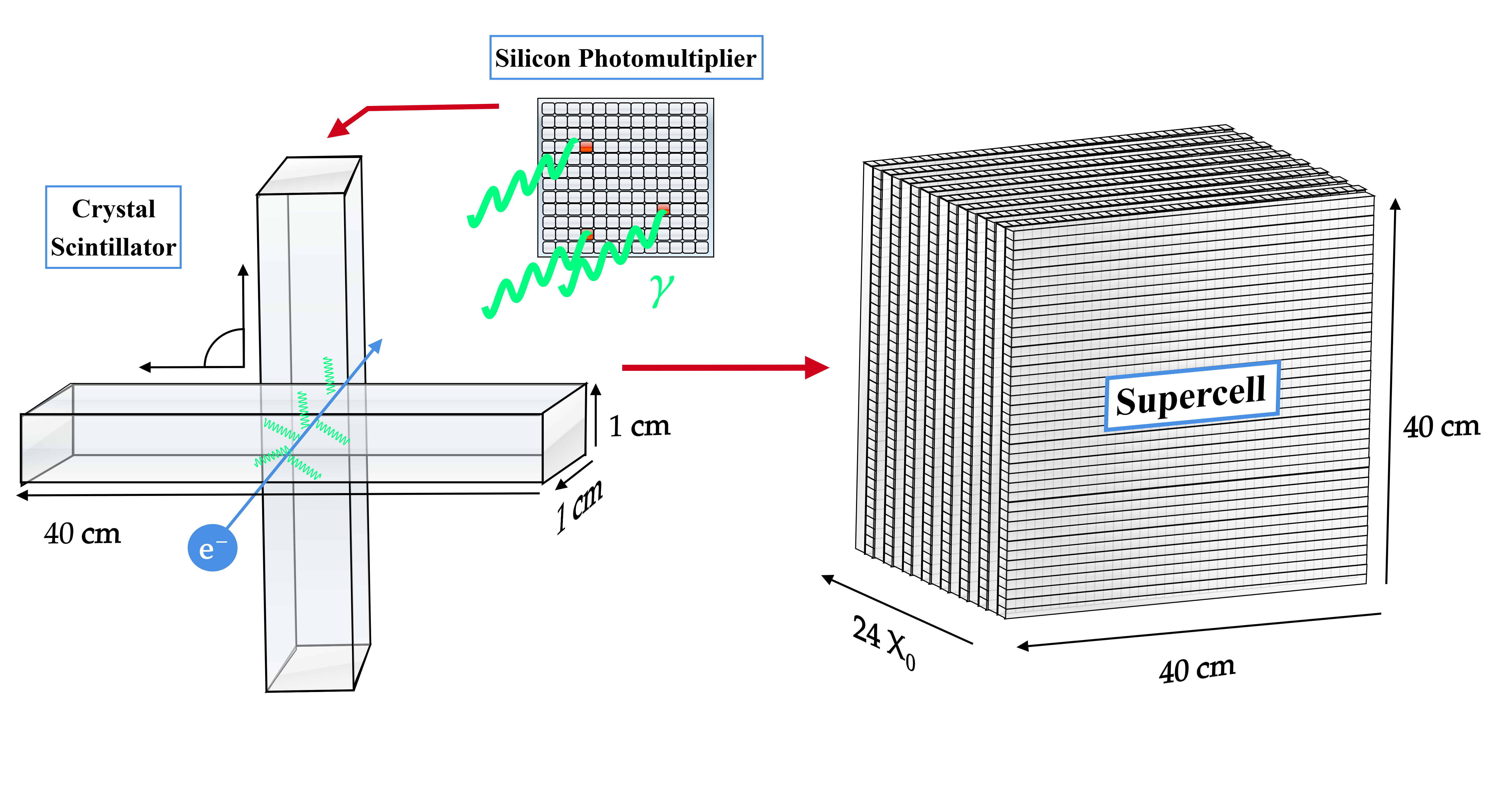}
    \caption{Diagram illustrating the design of the calorimeter.}
    \label{fig:fig_calorimeter}
\end{figure}

Further information can be found in Ref.~\cite{qi_rd_2022}. Fig.~\ref{fig:fig_calorimeter} illustrates the calorimeter design.

\subsubsection{Electromagnetic Shower Simulation}

A \gf simulation of electron showers is used in the study to evaluate the influence of dark noise on the measurement. The electrons had energies of 1,  2,  3,  4,  5,  7, 10, 15, 20, 25, 30, 40 and \qty{50}{\giga \electronvolt}, each with \num{1e4} events per step in energy.

\begin{figure}

\subfloat[\label{fig:fig_eventdisplay_1gev}]{\includegraphics[width=0.49\linewidth]{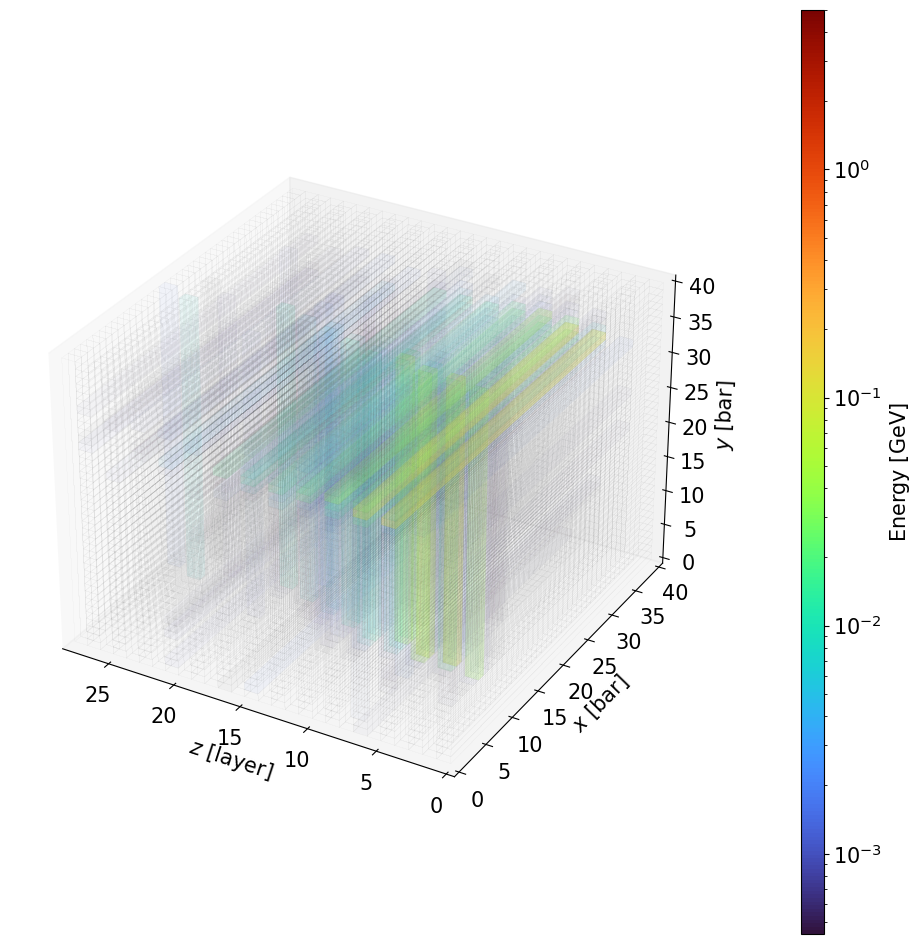}}
\hfill
\subfloat[ \label{fig:fig_eventdisplay_50gev}] {\includegraphics[width=0.49\linewidth]{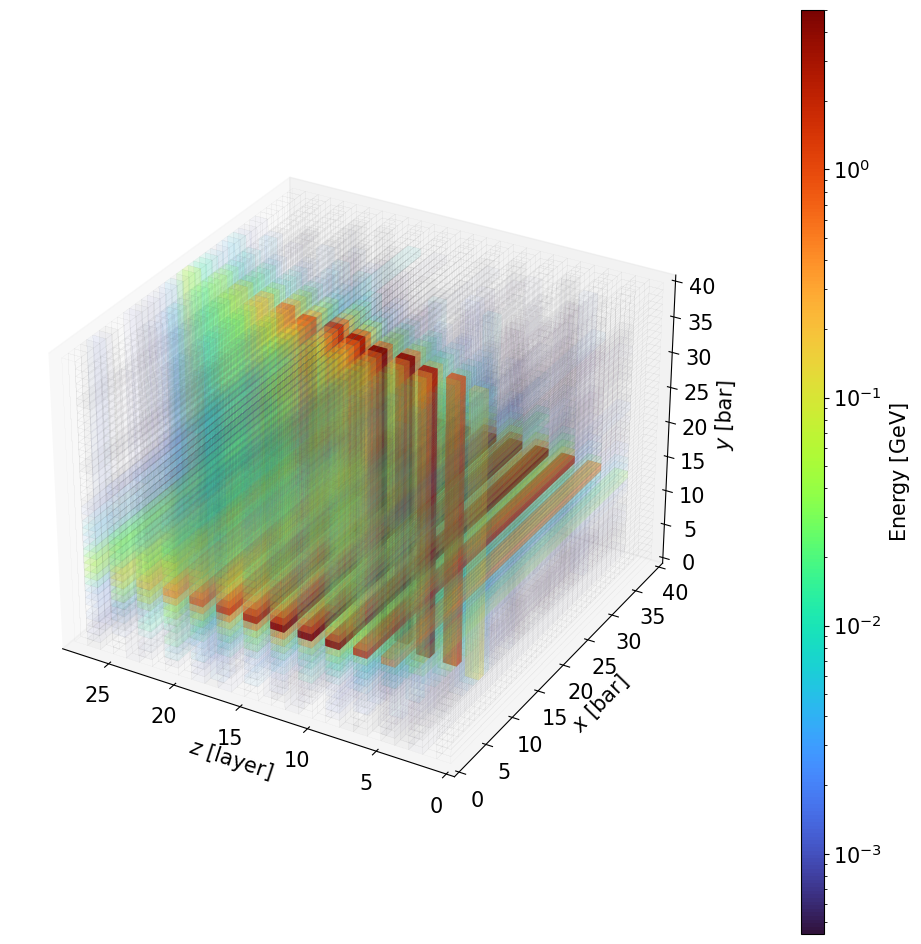}}

\caption{Event displays of a \qty{1}{\giga \electronvolt} (left) and \qty{50}{\giga \electronvolt} (right) shower deposited in the crystal calorimeter module. The colour (blue to red) and transparency (clear to opaque) indicate the energy deposited in the event, shown on the colour axis. The $x$ and $y$-axes indicate the index of the bar where the energy was deposited. The $z$-axis indicates the layer in which the energy was deposited.}

\label{fig:fig_eventdisplay}

\end{figure}

The raw simulation is digitised to model experimental and detector effects. Firstly, the digitisation accounts for scintillation effects: the Poisson fluctuations of the number of scintillation and attenuation of the light reaching the SiPMS due to self-absorption in the crystal and the photon detection efficiency (PDE) of the sensors. The effective light yield has been determined as \qty{200}{\mathrm{p.e.} \per \mathrm{MIP}} ($\qty{200}{\mathrm{p.e}} / \qty{8.9}{\mega \electronvolt}$). Secondly, SiPM effects such as saturation and cross-talk are considered utilising a method presented in Ref.~\cite{zhao_dynamic_2024}. Lastly, the effect of the readout electronics is considered, which affects the conversion of the SiPM charge to analogue-to-digital (ADC) channels. Three gain levels are used to obtain a broad dynamic range of the sensors. The threshold between low and medium/high gain is \qty{8e3}{\mathrm{p.e}}. The signal is also convolved with electronics noise as a normal distribution centred at  \qty{0}{\mathrm{p.e.}}  with an experimentally-obtained scale of \qty{1.56}{\mathrm{p.e.}}. The simulation was then calibrated to units of \unit{\giga \electronvolt}. The simulation did not include temperature or the effect of dark counts from the SiPM.  No noise threshold was included at this stage so that the influence of dark noise could be studied. Two event displays depicting a \qty{1}{\giga \electronvolt} and \qty{50}{\giga \electronvolt} shower, are shown in Fig.~\ref{fig:fig_eventdisplay}.

In the subsequent studies, energy is reverted into charge using the reciprocal of the average light yield. In practice, a noise energy threshold for each sensor is mandatory for the crystal ECAL to suppress not only dark noise as discussed in this paper, but also effects such as beam-induced background. A \qty{10}{\mathrm{p.e}} cut is proposed to be applied for this purpose, which is referred to as a noise cut henceforth. 

\begin{figure}[htpb]
    \centering
    \includegraphics[width=0.49\linewidth]{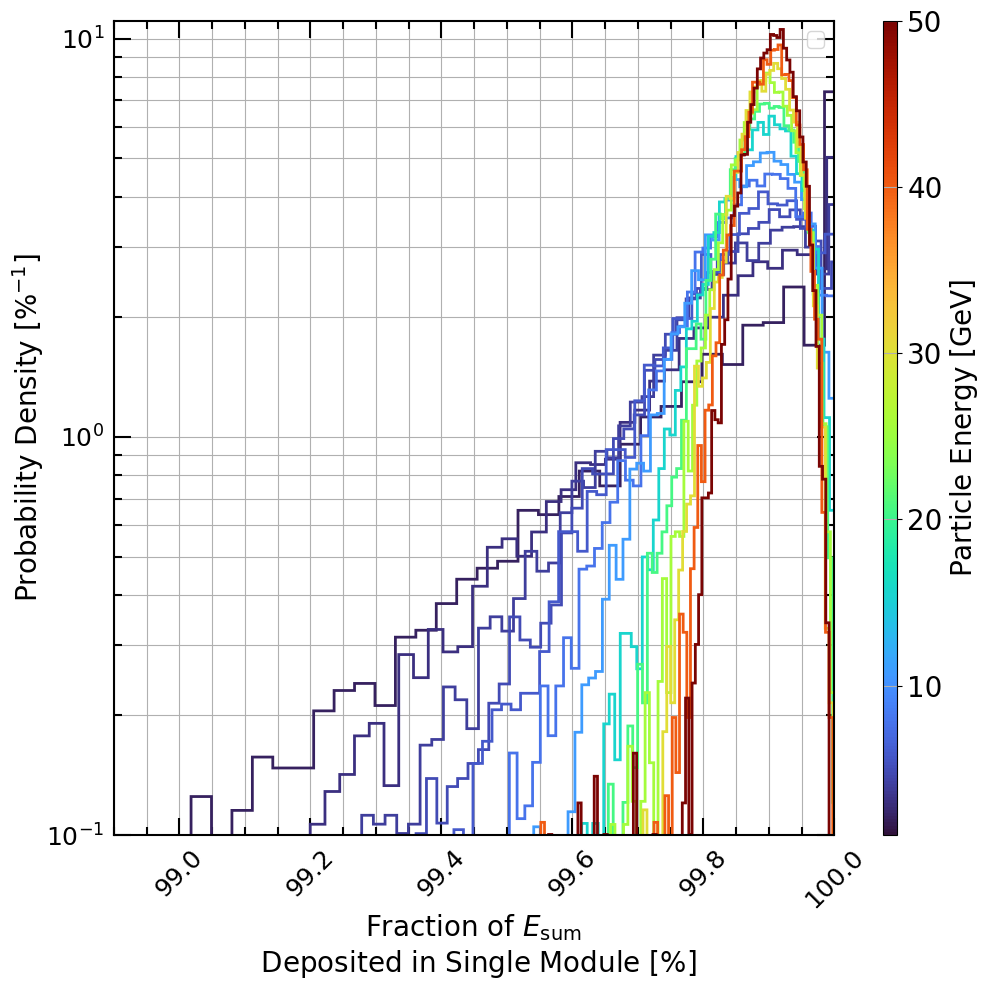}
    \caption{Distribution of the fraction of the total energy of the showers deposited in the module with the greatest number of active sensors. The colour scaling from blue to red indicates the particle energy.}
    \label{fig:fig_modulefraction}
\end{figure}

Most of the energy of the showers is deposited within a single module (supercell) of the calorimeter. This is demonstrated in Fig.~\ref{fig:fig_modulefraction}, which shows the ratio of energy deposited in the module with the greatest number of active sensors during an event to the total energy. It demonstrates that for showers in the range \qtyrange{1}{50}{\giga \electronvolt}, no less than \qty{99}{\percent} of the energy is deposited in a single module, and that the most probable value for the energy contained is close to \qty{99.9}{\percent}. This means that it is appropriate to consider only the \numproduct{2 x 1120} SiPMs belonging to one module as contributing noise to the energy measurement of a single shower.

\subsubsection{SiPM Dark Noise Simulation}

\begin{figure}[htpb]     \centering     \includegraphics[width=0.49\linewidth]{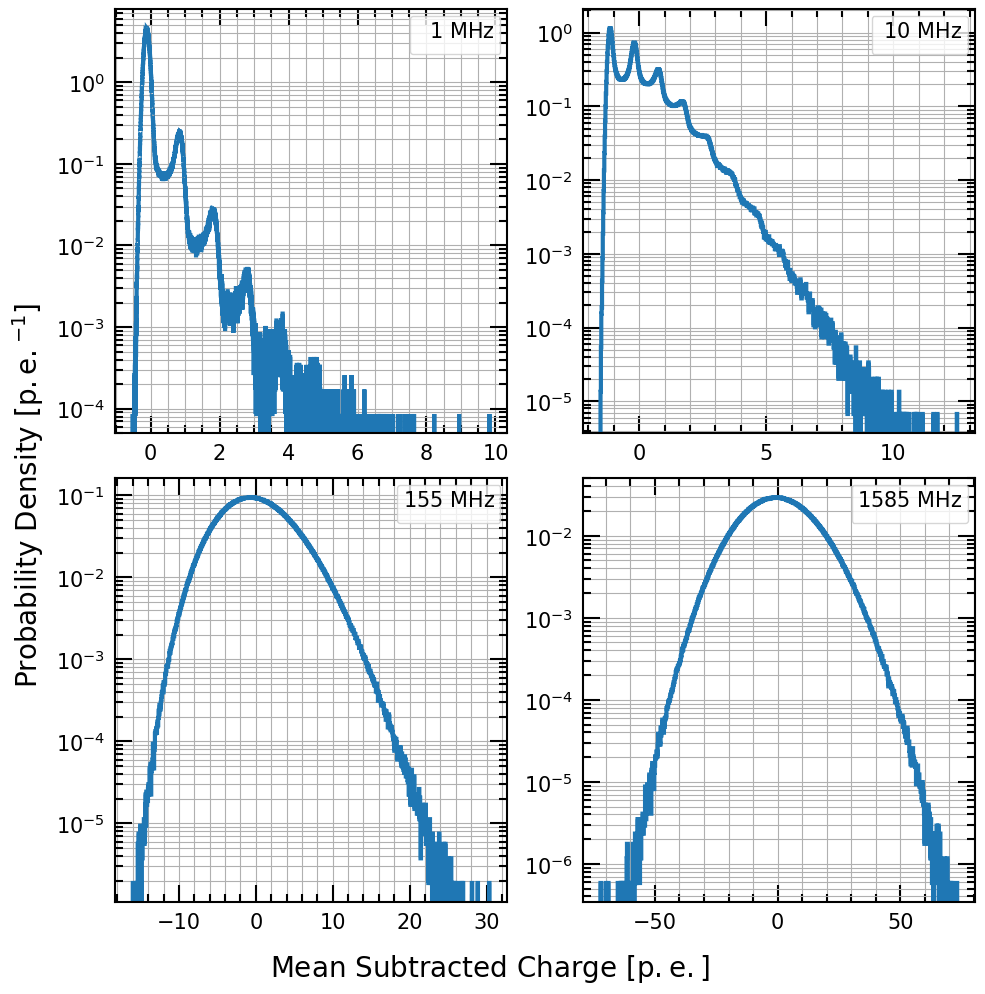}     
\caption{Distributions of simulated dark noise distributions with the mean subtracted from the distribution. The x-axis shows the charge in the number of photoelectrons (\unit{\mathrm{p.e}}). The y-axis represents the probability density. The DCR of the simulation is shown in units  \unit{\mega \hertz}.  }    
\label{fig:fig_darksim}
\end{figure}

A flexible simulation program for SiPM charge spectra was developed in Ref.~\cite{garutti_simulation_2021}. This program allows for the charge spectra of SiPMs to be simulated with several options for distributions of photon number and time, as well as various models to describe experimental noise effects, such as after-pulsing from de-trapping of charge carriers in the silicon and dark noise. This model operates by simulating the amplitude, number, and time of SiPM pulses from different sources and determining the total charge measured by the SiPM from the times at which the pulse occurs, the shape of the pulse, and the length of the integration window (gate) for the pulse. The simulation program outputs the charge spectrum in units of the number of photoelectrons (\unit{\mathrm{p.e.}}).  This model describes the number of dark discharges that occur during the period $t_{0}$ to $t_{\mathrm{gate}}$, $t_{0}$ is the lower limit of the time at which dark pulses are considered before the gate and $t_{\mathrm{gate}}$ is the length of the integration gate for the pulse. The number of dark counts is described using a Poisson distribution characterised by $\mu_{\mathrm{dark}} = DCR \cdot (t_{\mathrm{gate}}-t_{0})$, where $DCR$ is the dark count rate. The time distribution of the pulses is given by a uniform distribution from $t_{0}$ to $t_{\mathrm{gate}}$. Further details of the pulse shape and charge model can be found in Ref.~\cite{garutti_simulation_2021}.  Simulations of \qty{1e7}{\mathrm{events}} were generated using the program at 20 logarithmically increasing values of DCR from \qty{1}{\mega \hertz}-\qty{1.58}{\giga \hertz}. The remaining parameters were held constant at nominal values for a \qtyproduct{3 x 3}{\milli \meter \squared} sensor. The parameters used are listed in the Appendix Table \ref{tab:SiPMSimulationParameters}. A set of mean-subtracted dark-noise distributions is shown in Fig. ~\ref{fig:fig_darksim}.  

The purpose of mean subtraction is to mitigate the effect of the increasing mean charge deposited owing to the noise. $\mu_{\mathrm{dark}}$ increases in fluence in proportion to the relationship shown in Appendix Fig.~\ref{fig:fig_dcrtofluence}, assuming constant gate length. This means the sums of dark noise and physics signal will be 'shifted' to ever higher charges with increasing fluence. To mitigate this effect, it is assumed the mean of the dark count distribution can be measured and subtracted from the charge measured by the SiPM.  This could be achieved in practice by triggering the readout randomly in the time window between the bunch crossings of a collider experiment (i.e. uncorrelated in time) to obtain a signal dominated by dark noise, which was then used to determine the mean. However, this method is speculative and is yet to be validated.  

\begin{figure}[htpb]  
\centering    
\includegraphics[width=0.49\linewidth]{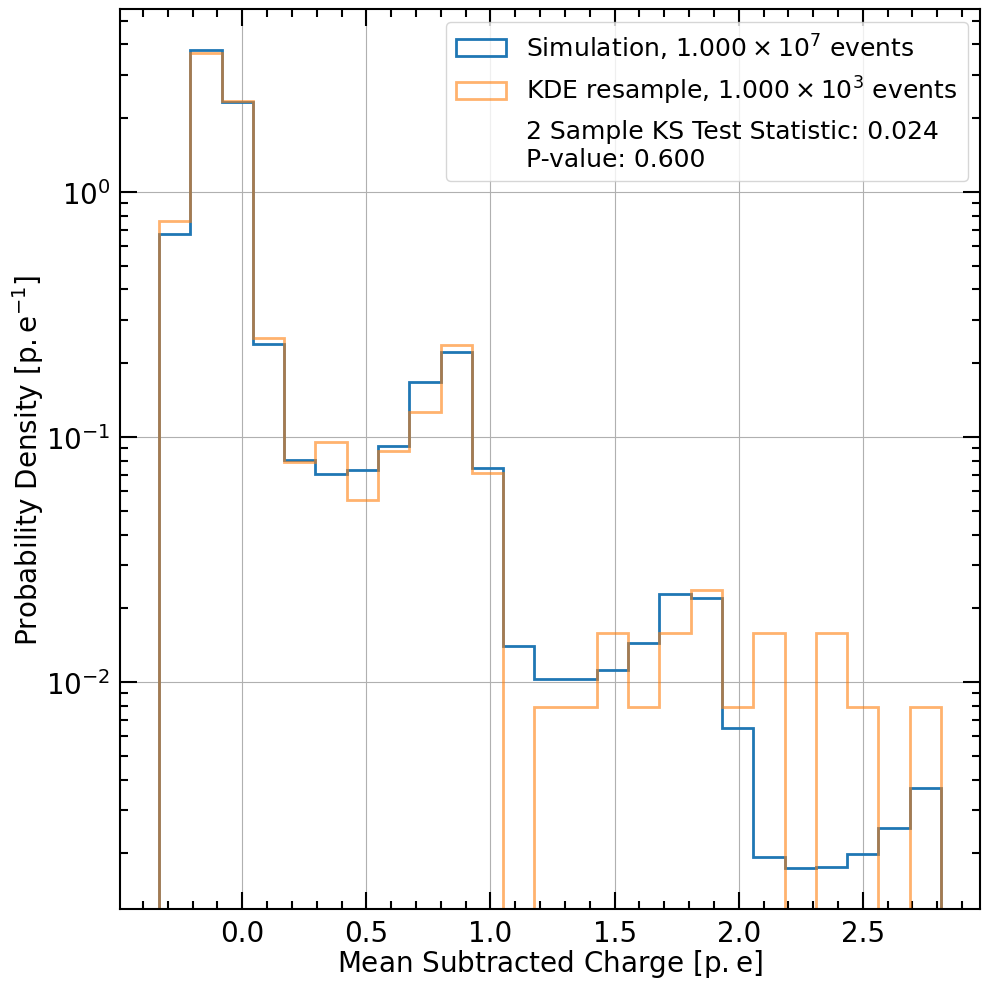}     
\caption{Distributions of a simulated dark noise distributions at \qty{1}{\mega \hertz} DCR in blue, and a KDE resample of \qty{1e3}{\mathrm{events}} shown in orange. The mean has been subtracted from the distribution. The KS-statistic and p-value of a 2-sample Kolgromov-Smirnov test are shown in the legend.}   
\label{fig:fig_kderesample}
\end{figure}

All sensors of the module produce dark noise depending on the level of irradiation. This means that a charge from the dark count distribution must be sampled for every sensor in the module event by event, which may then be added to the signal to study its influence. However, it is not feasible to simulate dark noise in terms of computation time and storage, particularly when $\mu_{\mathrm{dark}} \gg 1$ and many individual pulses must be considered. Therefore, a kernel density estimate (KDE) was used to generate resamples from a simulation produced by the program. A KDE describes each data point by a distribution centred around that point \cite{rosenblatt_remarks_1956, parzen_estimation_1962}. The most commonly used distribution is the normal distribution. The width of the distribution ('bandwidth') controls the smoothing of the estimate and can be determined via an algorithm. In this study, a standard bandwidth selection method was used \cite{sheather_reliable_1991}. The simulation can then be resampled by randomly selecting charges from the simulation produced by the program, with replacement and perturbing the value by normally distributed noise scaled to the bandwidth. An example histogram comparing a resample of \qty{1e3}{\mathrm{events}} to the original simulation is shown in Fig. ~\ref{fig:fig_kderesample}. The results of the  2-sample Kolgromov-Smirnov test are included in the legend. The $p$-value indicates that the null hypothesis cannot be rejected with \qty{99}{\percent} confidence ($p < 0.01$). This result suggests that both samples were likely to be drawn from the same distribution, demonstrating the effectiveness of this resampling technique.

\subsection{Analysis}
The analysis procedure is outlined henceforth. 

For each event of the electron shower simulation, the module where the most cells were active was found. Then, a cut was applied to remove cells from other modules that contributed energy to the event. As shown in Fig.~\ref{fig:fig_modulefraction}, almost all the energy from a single event is deposited in a single module. Subsequently, the cut has a negligible effect on the reconstructed energy of the electron. The physics signal was converted from energy in \unit{\giga \electronvolt} to charge in \unit{{\mathrm{p.e}}} according to the mean light yield of the scintillating crystal ($\qty{200}{\mathrm{p.e}} / \qty{8.9}{\mega \electronvolt}$). If fewer than 2240 cells were active during the event, the remaining cells in the module were considered to have measured \qty{0}{\mathrm{p.e}} charge.  

The resampling process was performed for each of the measurements discussed henceforth.  For each event and each step in DCR, 2240 dark counts were resampled from the SiPM dark noise distribution and divided into two sets of 1120 counts for the two SiPMs for each crystal bar, respectively. If the total number of active cells in the event were less than the number in the module, those cells that did not register a signal were treated as having measured \qty{0}{\mathrm{p.e}}. The dark noise and physics signal were then added to obtain the signal that would have been measured at a particular step in DCR. 

Firstly, the effect of the noise on the response of the individual sensors was studied. A subsample of \qty{1000}{\mathrm{events}} was selected from the sample of \qty{1}{\giga \electronvolt} electron showers. The charges measured by the sensor at each step in DCR were then recorded, with and without the application of a \qty{10}{\mathrm{p.e.}} noise cut. The average charge measured in each cell was calculated for each step in DCR for both cases. 

Secondly, using all the available energies and events of electron showers, the mean false positive rate of the noise cut trigger was measured. Assuming a perfect trigger, the false positive rate is defined in Eq.~\ref{eq:eq_FPR}:

\begin{equation}
    FPR = \frac{FP}{TN + FP} = \frac{\#(Q_{\mathrm{dark}} + Q_{\mathrm{signal}} \geq C \mid Q_{\mathrm{signal} }< C)}{\#(Q_{\mathrm{signal}} < C)}
    \label{eq:eq_FPR}
\end{equation}

where $\#$ indicates a count, $Q_\mathrm{dark}$ and $Q_{\mathrm{signal}}$ indicates the charge measured by the SiPMs from dark counts and physics signal, respectively, $C$ indicates the noise threshold of \qty{10}{\mathrm{p.e}} and and $\mid$ represents the conditional bar, which denotes that the count is conditioned on the event that $Q_{\mathrm{signal}}$ < C. Explicitly, Eq.~\ref{eq:eq_FPR} represents the probability that a charge that should have been rejected by the cut satisfies it due to the presence of dark noise. 

Lastly, using all the available energies and events of electron showers, the sum of the charge of all the sensors of the module ($E_{\mathrm{sum}}$) was calculated at each step in DCR. The charge was then converted back to units of energy using the reciprocal of the light yield conversion to obtain the measured calorimeter response to the electron at a particular step in DCR. The distributions of energy were studied both with and without a \qty{10}{\mathrm{p.e.}} cut applied. Then,  for each particle energy and DCR step, the distributions of $E_{\mathrm{sum}}$ were fitted using an extended unbinned log-likelihood fit using a Crystal Ball distribution, defined according to Eq.~\ref{eq:eq_crystal_ball}  \cite{gaiser_charmonium_1982}: 

\begin{equation}
    f(z, \beta, m) = 
\begin{cases} 
    N \exp\left(-\frac{z^2}{2}\right), & \text{for } z > -\beta \\ 
    N A (B - z)^{-m}, & \text{for } z \leq -\beta 
\end{cases}
\label{eq:eq_crystal_ball}
\end{equation}

Where: $A = \left(\frac{m}{|\beta|}\right)^{m} \exp\left(-\frac{\beta^2}{2}\right)$, $B = \frac{m}{|\beta|} - |\beta|$,
$N$ is a normalisation constant, and $z$ is the value of $E_{\mathrm{sum}}$ normalised to the mean ($\mu$) and standard deviation ($\sigma$), which are free parameters in the fit. This distribution is used due to a power-law left-sided tail, which can describe leakage. Leakage in this design of ECAL is predominantly from albedo, rather than longitudinal leakage. 

The linearity of the calorimeter was measured by comparing the value of $\mu$ to the known particle energy in simulation. The percentage deviation from this value was used to quantify any bias.

The resolution was measured using the fitted values of $\mu$ and $\sigma$, the resolution is estimated using $R = \frac{\mu}{\sigma}$. The resolution of the calorimeter is then fitted using least-squares regression for each step in DCR as a function of particle energy with Eq.~\ref{eq:eq_resolution} \cite{wigmans_calorimetry_2017}:

\begin{equation}
    R = \frac{s}{\sqrt{E}} \oplus c \oplus \frac{n}{E}
    \label{eq:eq_resolution}
\end{equation}

where $E$ is the particle energy, $s$ is the stochastic term describing the influence of shower fluctuations on the resolution, $c$ is a constant term describing the calibration quality and uniformity and $n$ is a noise term. The influence of the dark noise on the calorimeter performance is then presented as a function of fluence. All fits were performed using the \texttt{MIGRAD} algorithm implemented in \iminuit \cite{dembinski_scikit-hepiminuit_2024, james_minuit_1975}. As a caveat, no other experimental effects than radiation-induced dark noise are considered in the measurement of the resolution, and therefore the resolutions presented in this study are significantly superior to experimental data. The relationship of Appendix Fig.~\ref{fig:fig_dcrtofluence} was used to relate DCR to fluence, under the assumption that the calorimeter SiPMs were operated at \qty{20}{\celsius}.  Details of how DCR is related to fluence using a study performed in Ref.~\cite{altamura_radiation_2023} at constant temperature is given in Appendix Sec.~\ref{sec:DCR_vs_fluence}. 

Finally, the temperature stability of the detector without irradiation is estimated using the average temperatures measured over a week of operation during a 2024 Testbeam at the Super Proton Synchrotron at CERN in Geneva, Switzerland, using the estimated DCR at \qty{0}{\per \centi \meter \squared} fluence. Detail of how DCR is related to temperature at a constant fluence using a study performed in Ref.~\cite{otte_characterization_2017} is given in Appendix Sec.~\ref{sec:DCR_vs_T}. 

All histograms are presented using a standard bin width optimisation method \cite{knuth_optimal_2019}.

\section{Results}

\subsection{Hit Energy Distribution}

\begin{figure}[htpb]
\subfloat[\label{fig:fig_hitenergysample}]{\includegraphics[width=0.49\linewidth]{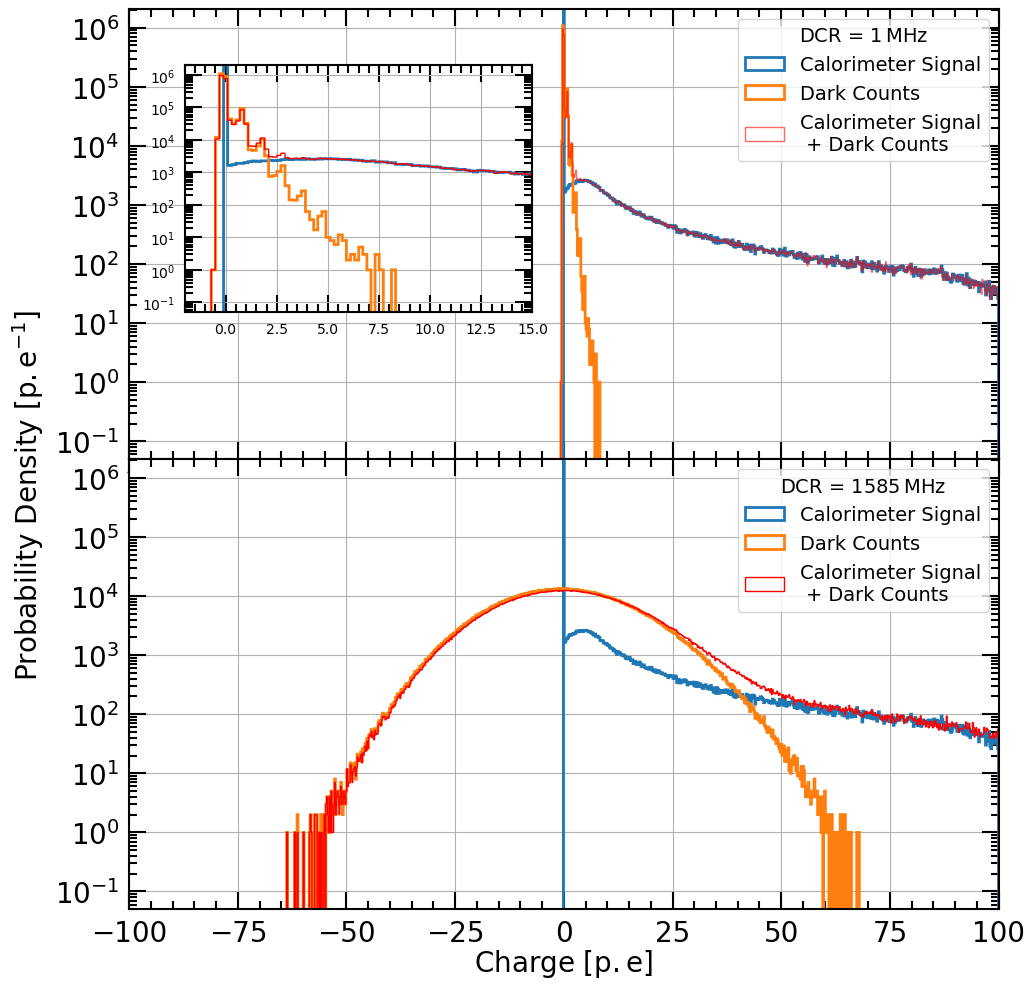}}
\subfloat[ \label{fig:fig_hitenergysamplecut}]   {\includegraphics[width=0.49\linewidth]{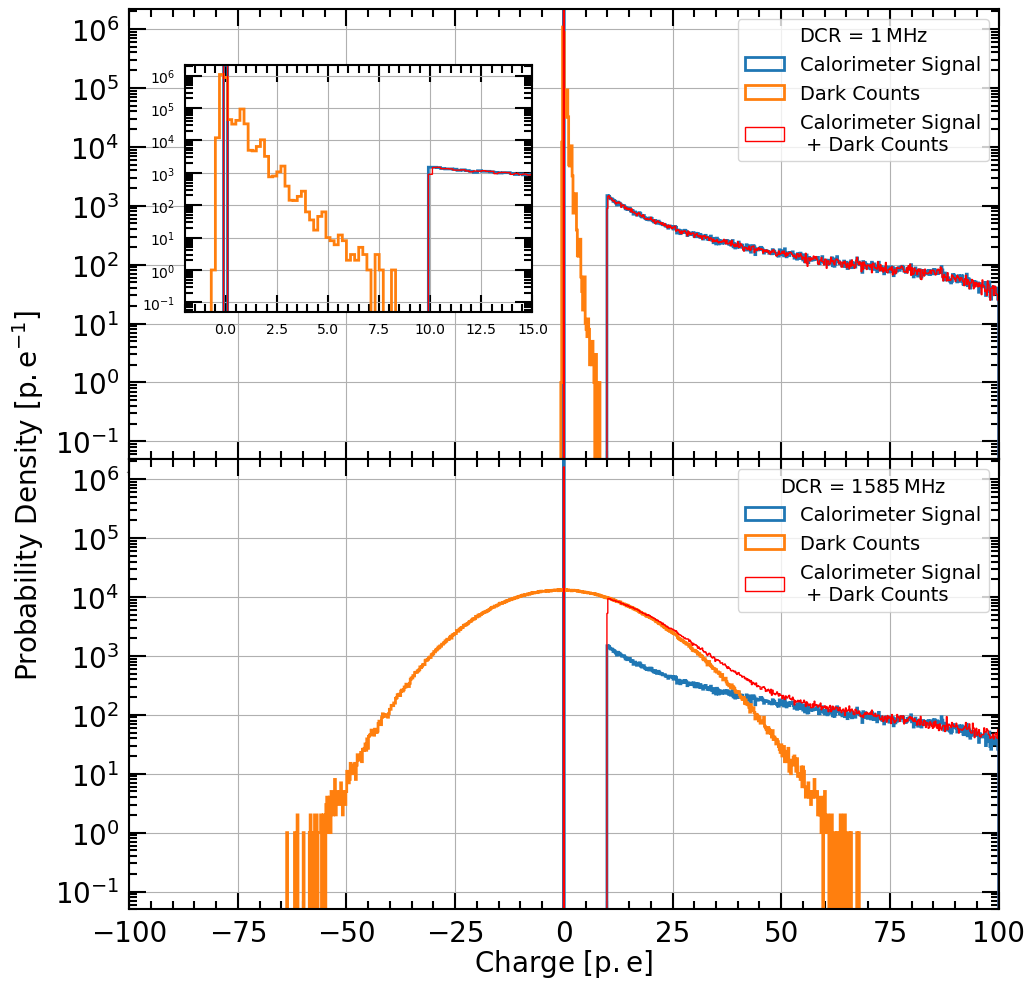}}
\hfill
 \caption{Distributions of charge measured for a \qty{1}{\giga \electronvolt} electron shower, at \qty{1}{\mega \hertz} (top) and \qty{1.585}{\giga \hertz} (\qty{1585}{\mega \hertz}, bottom) DCR. The distribution is shown without (left) and with (right) the cut applied. The blue line indicates the distribution of the energy deposited in a module, without noise. The orange line shows the distribution of noise. The red line shows the distribution of their sum. The upper left inset shows the \qty{1}{\mega \hertz} sample, zoomed to $Q < \qty{15}{\mathrm{p.e.}}$ }
\label{fig:fig_hitenergy}
\end{figure}

Fig. ~\ref{fig:fig_hitenergy} shows the cell charge distribution and how it is affected by noise at the extreme values of DCR in the sample, with and without the application of a \qty{10}{\mathrm{p.e.}} noise cut. The top plots of Figs.~\ref{fig:fig_hitenergysample} and \ref{fig:fig_hitenergysamplecut} show that the cut is sufficient to be able to virtually eliminate the contribution of noise from sensors which did not measure a physics signal at a nominal DCR level of \qty{1}{\mega \hertz}. However, the bottom plot shows that at extreme values of DCR, the dark noise is at a sufficiently high level that the limit is no longer an effective means of excluding its contribution.

\begin{figure}
    \centering
    \includegraphics[width=0.49\linewidth]{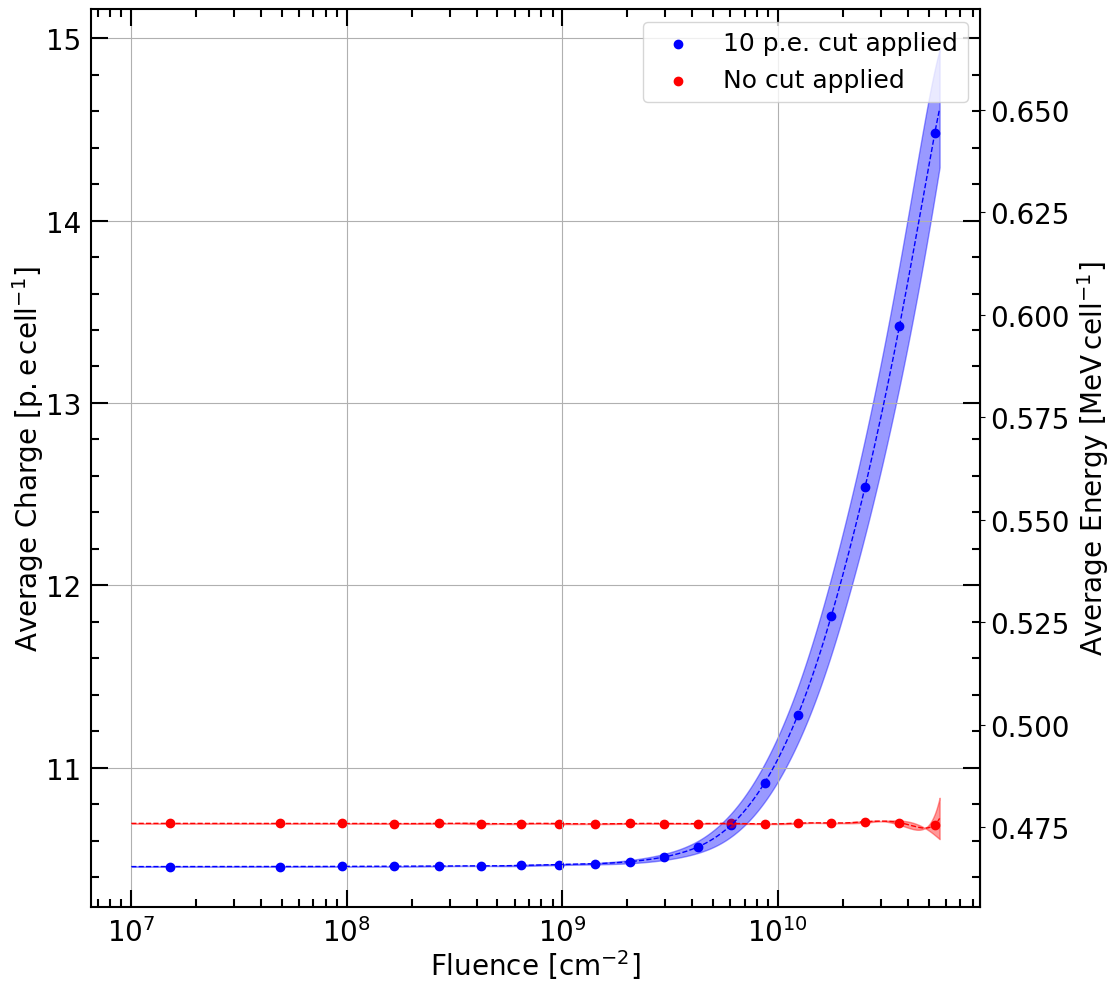}
    \caption{Average charge (energy) measured by each sensor as a function of fluence. The blue and red dots indicate the average charge (energy) measured by the sensor, with and without a \qty{10}{\mathrm{p.e.}} noise cut applied. The dashed lines and shaded regions indicate the spline fit used to obtain the fluence and its 1-$\sigma$ confidence band. Values with $DCR$ outside the valid range in Fig.~\ref{fig:fig_dcrtofluence} cannot be shown.}
    \label{fig:fig_hitenergyshift}
\end{figure}

Fig.~\ref{fig:fig_hitenergyshift} illustrates the effect of the additional smearing caused by the noise. Since there is only minor asymmetry of the dark count distribution about its mean value in all cases, the mean of the hit energy distribution is not considerably affected by the noise if no cut is applied. However, when a cut is enforced, the hit energy distribution becomes biased to the right tail of the noise distribution. In other words, when the noise level is sufficient that it can considerably surpass the trigger condition, the sensor experiences a considerable bias. At the maximum value of fluence studied it is around \qty{0.65}{\mega \electronvolt} per sensor.

In summary, the presented results indicate that dark noise will impose a bias on the measured signal if a noise cut is imposed. It is expected that this effect will, in turn, bias the total shower energy measured by the calorimeter at fluences greater than \qty{1e10}{\per \centi \meter \squared}.

\subsection{Energy Sum Distribution}

\begin{figure}[htpb]
\subfloat[\label{fig:fig_energysums_nocut_1gev}]{\includegraphics[width=0.49\linewidth]{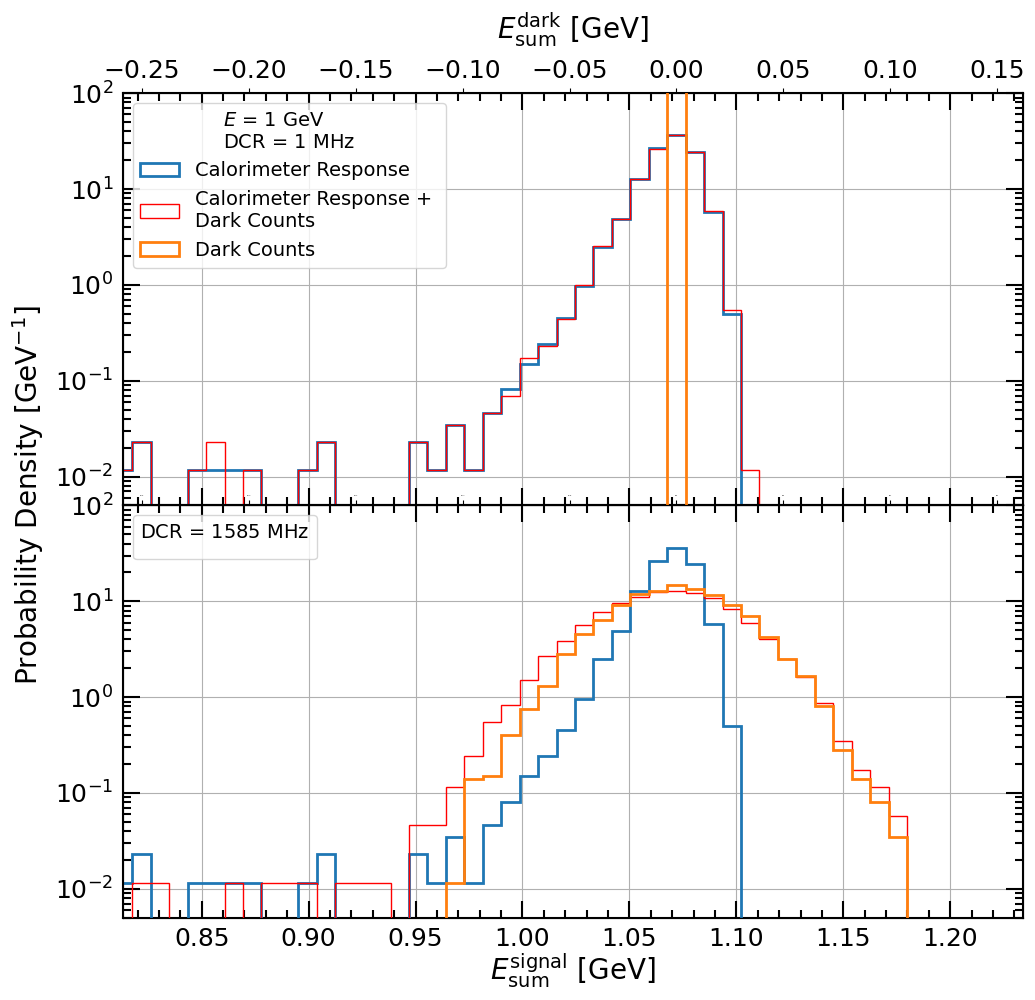}}
\subfloat[ \label{fig:fig_energysums_nocut_50gev}]   {\includegraphics[width=0.49\linewidth]{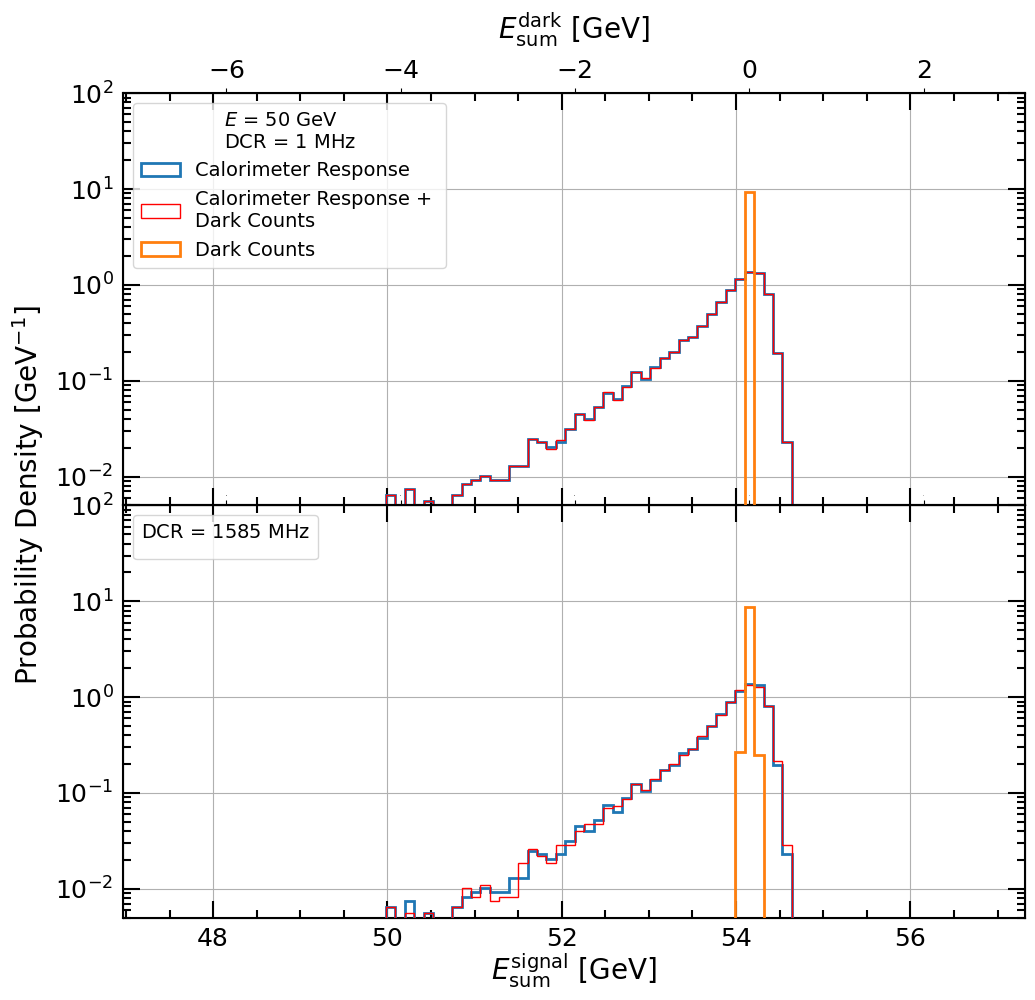}}
\hfill
\caption{Energy distributions for a \qty{1}{\giga \electronvolt} electron shower (left) and another \qty{1}{\giga \electronvolt} shower (right) without a noise cut applied. The blue line shows the calorimeter response without dark noise, while the red line shows the combined signal and dark noise. The bottom x-axis indicates the units for these distributions. The orange line represents energy from dark noise alone, with units shown on the top x-axis.}
\label{fig:fig_energysums_nocut}
\end{figure}

\begin{figure}
\subfloat[\label{fig:fig_energysums_cut_1gev}]{\includegraphics[width=0.49\linewidth]{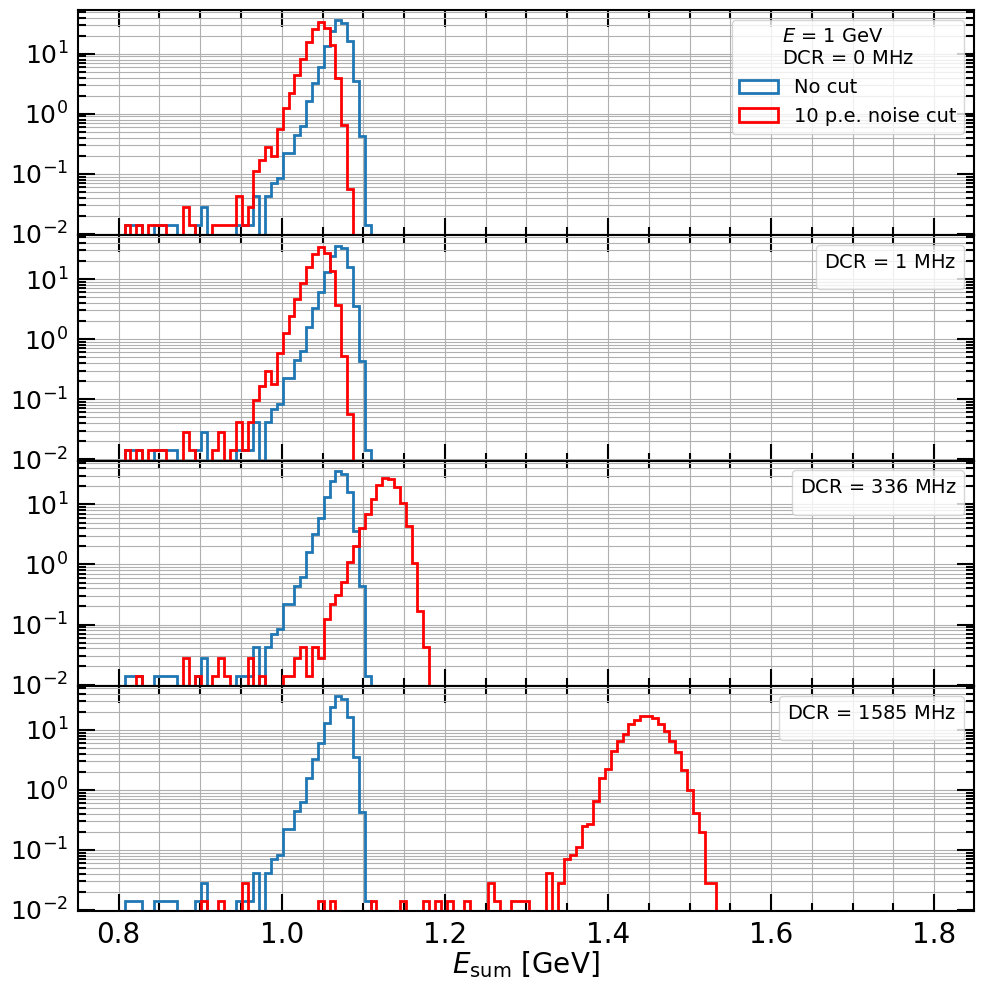}}
\subfloat[ \label{fig:fig_energysums_cut_50gev}]   {\includegraphics[width=0.49\linewidth]{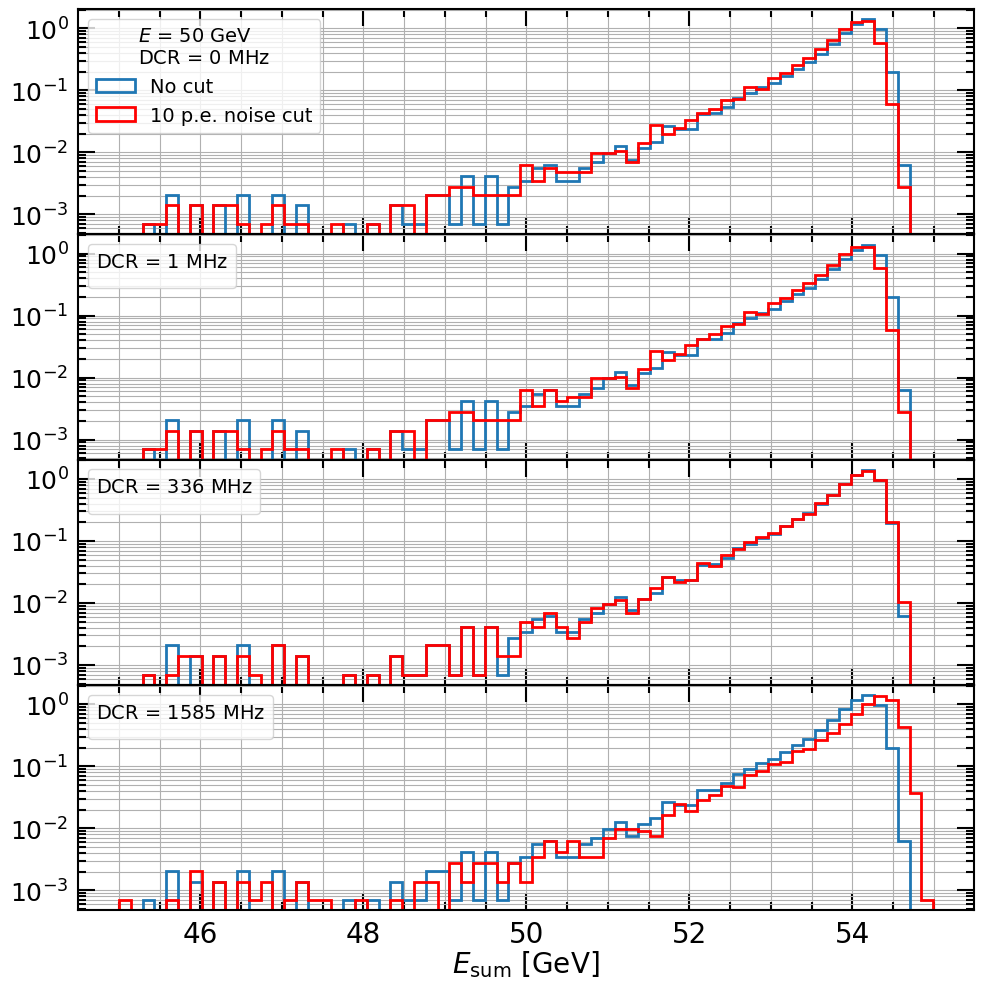}}
\hfill
\caption{Energy distributions for a \qty{1}{\giga \electronvolt} electron showers (left) and \qty{50}{\giga \electronvolt} electron showers (right) with a \qty{10}{\text{p.e.}} noise cut applied. The blue line shows the calorimeter response without dark noise, while the red line shows the combined signal and dark noise. Different levels of DCR are indicated in the legends.}
\label{fig:fig_energysums_cut}
\end{figure}

Fig.~\ref{fig:fig_energysums_nocut} shows energy distributions for electron showers with energies of \qty{1}{\giga \electronvolt} and \qty{50}{\giga \electronvolt}, without a noise cut applied. In the \qty{1}{\giga \electronvolt} case shown in Fig.~ \ref{fig:fig_energysums_nocut_1gev}, noise significantly smears the signal at high DCR. By contrast, the \qty{50}{\giga \electronvolt} shower energy (Figure \ref{fig:fig_energysums_nocut_50gev}) is essentially unaffected by noise. This can be explained by the fact that the distribution of the sum of dark noise has a variance that is independent of particle energy and depends only on the DCR, while the variance of the calorimeter's response to the electron increases with the energy of the incident particle. Therefore, low-energy showers are expected to experience greater fluctuations in reconstructed energy compared to high-energy showers compared to their mean due to dark noise. Considerable non-linearity of the response of \qtyrange{5}{8}{\percent} can be seen in the figures. This discrepancy is caused by SiPM cross-talk in the digitisation procedure in the range and is not observed if digitisation is not applied.

Fig.~\ref{fig:fig_energysums_cut} shows energy distributions for electron showers with energies of \qty{1}{\giga \electronvolt} and \qty{50}{\giga \electronvolt}, with a \qty{10}{\mathrm{p.e.}} noise cut applied. As expected from Fig.~\ref{fig:fig_hitenergyshift}, a systematic 'shift' in the energy is observed as DCR increases. Furthermore, as indicated by Fig.~\ref{fig:fig_hitenergyshift}, the cut will result in a smaller calorimeter signal below around \qty{1e10}{\per \centi \meter \squared} fluence and rapidly increase at higher fluences. This is observed for both Fig.~\ref{fig:fig_energysums_cut_1gev} and Fig.~\ref{fig:fig_energysums_cut_50gev}. Lower shower energies are affected more greatly than higher energies, which is again due to the independence of the distribution of the sum of dark noise with particle energy. 

In summary, the presented results indicate that the effect of noise is energy-independent, and therefore degrades the calorimeter performance for low-energy showers more strongly than for high-energy ones. Additionally, the addition of a \qty{10}{\mathrm{p.e}} noise cut results in an energy bias, which is explained by the observations made in Fig.~\ref{fig:fig_hitenergyshift}.

\subsection{False Positive Rate of Cut Trigger}

\begin{figure}
    \centering
    \includegraphics[width=0.49\linewidth]{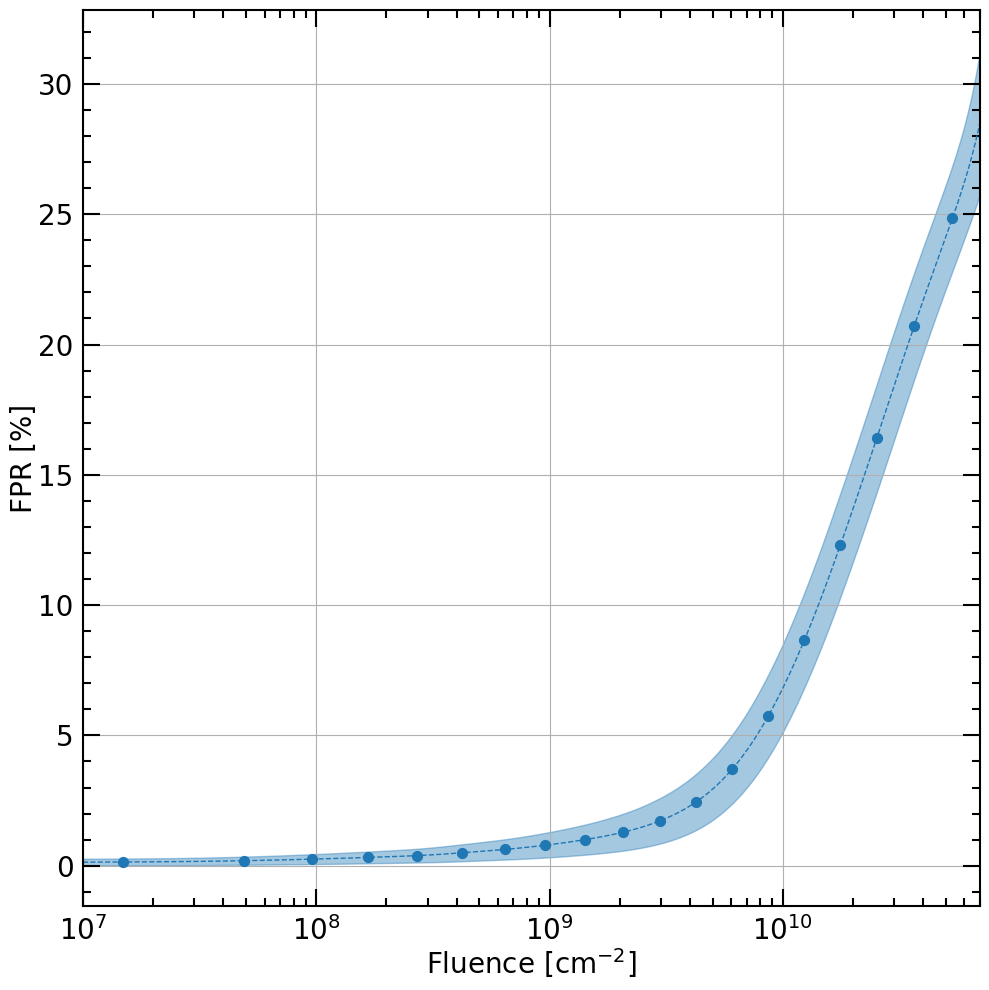}
    \caption{The false positive rate ($FPR$) of the \qty{10}{\mathrm{p.e}} noise cut versus fluence, defined according to Eq.~\ref{eq:eq_FPR}. The blue points indicate the $FPR$. The dashed lines and shaded regions indicate the spline fit used to obtain the fluence and its 1-$\sigma$ confidence band. Values with $DCR$ outside the valid range in Fig.~\ref{fig:fig_dcrtofluence} cannot be shown.}
    \label{fig:fig_fpr}
\end{figure}

The false positive rate of the trigger is shown as a function of fluence in Fig.~\ref{fig:fig_fpr}. In summary, it shows that the false positive rate gradually increases above fluences of \qty{1e8}{\per \centi \meter \squared}. The $FPR$ surpasses \qty{5}{\percent} at a fluence of around \qty{1e10}{\per \centi \meter \squared} fluence and increases past that value. This result indicates the diminishing effectiveness of the trigger at distinguishing signal from noise.

\subsection{Resolution And Linearity}

\begin{figure}[htpb]
    \centering
    \includegraphics[width=0.8\linewidth]{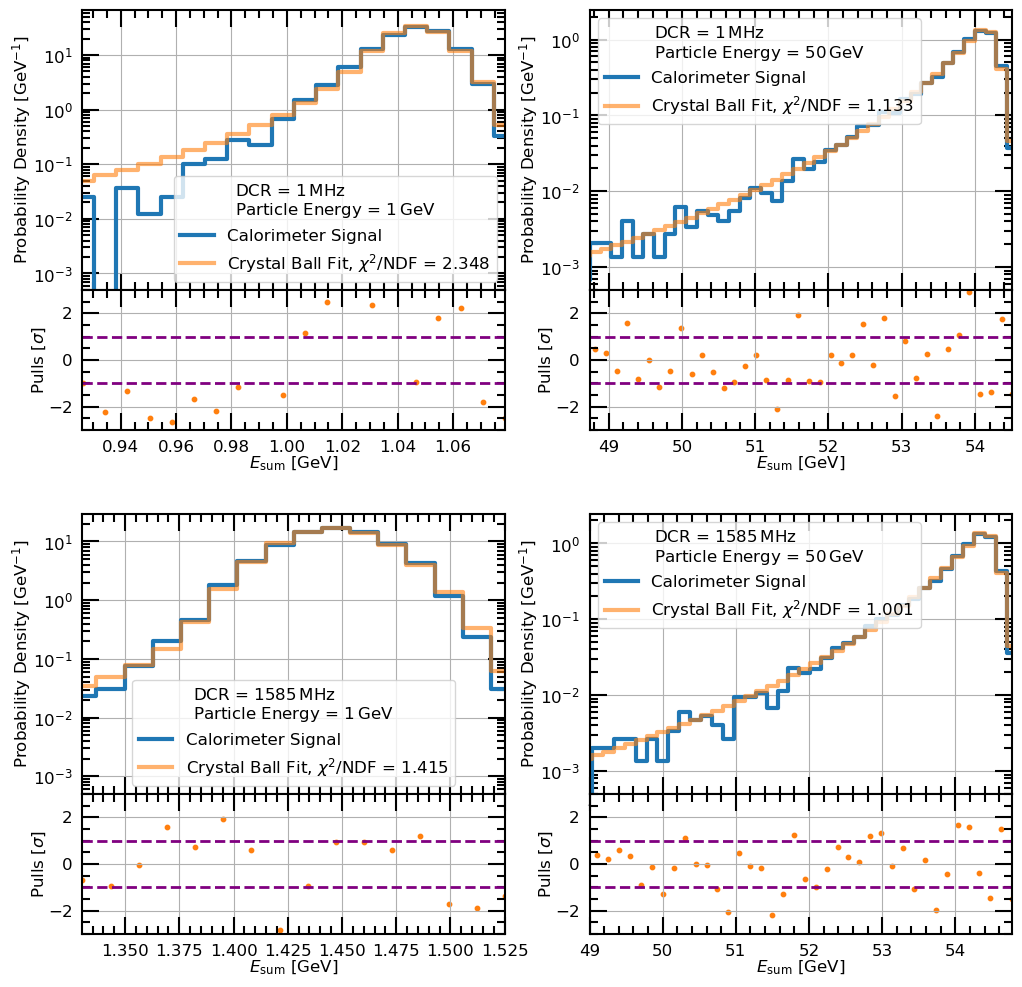}
    \caption{The energy sum distribution of simulation, shown in blue, fitted with the Crystal Ball distribution, shown in orange. The top (bottom) subplots show the lowest (highest) DCR step. The left (right) plots show the \qty{1}{\giga \electronvolt} (\qty{50}{\giga \electronvolt}) samples. The pulls of the fit are shown at the bottom of each figure as orange points. The fit and simulation were binned using a standard bin width optimisation method \cite{knuth_optimal_2019}. }
    \label{fig:fig_crystalball_fit}
\end{figure}

Fits of the Crystal Ball distribution to the reconstructed energy sum are shown in Fig.~\ref{fig:fig_crystalball_fit} and compared to the fit of Eq.~\ref{eq:eq_crystal_ball} by integrating over the bins.  Overall, the reduced $\chi^{2}$ of all fits as a function of DCR and particle energy was found to be $\frac{\chi^{2}}{\mathrm{NDF}} = \num{1.36 \pm 0.52}$. This result indicates an acceptable fit of the model to the simulation, even in the presence of dark noise.

\begin{figure}[htpb]
\subfloat[\label{fig:fig_linearity}]{\includegraphics[width=0.49\linewidth]{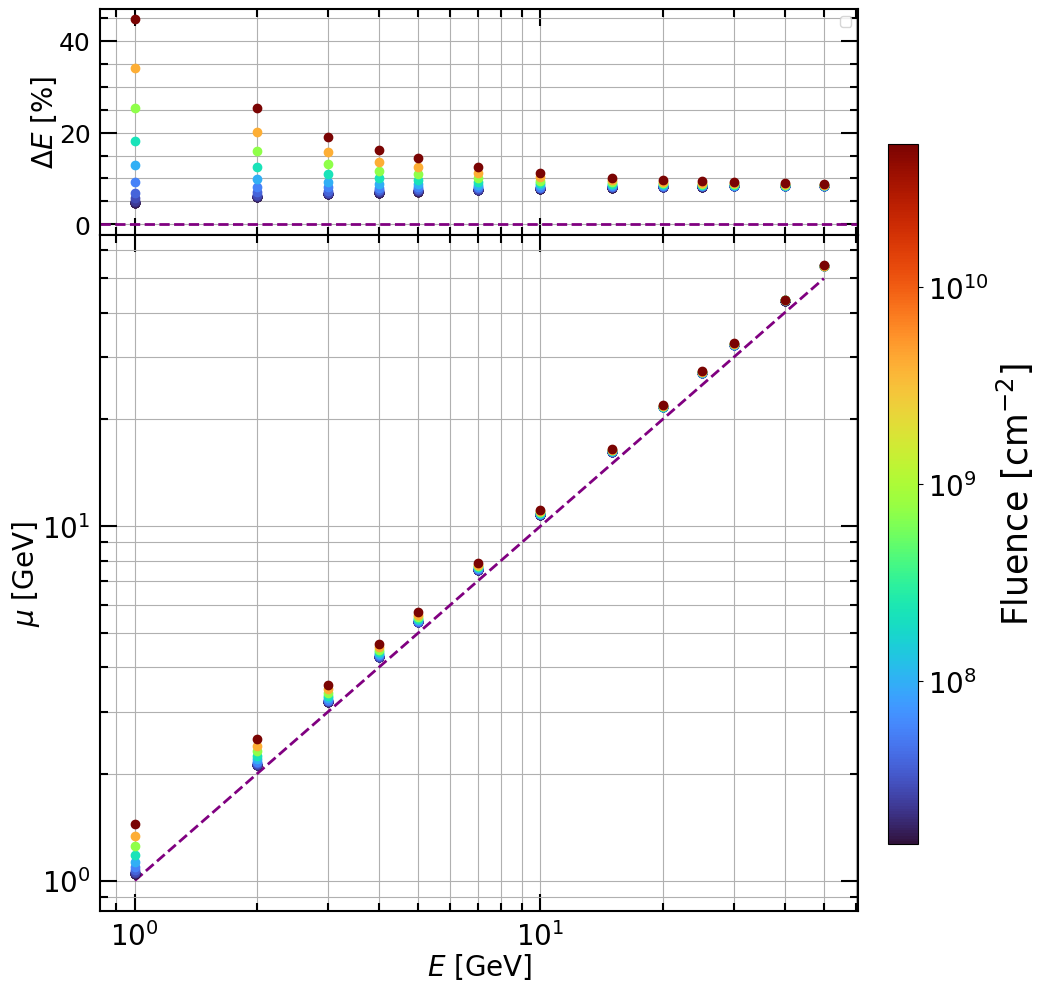}}
\subfloat[ \label{fig:fig_resolution}]   {\includegraphics[width=0.49\linewidth]{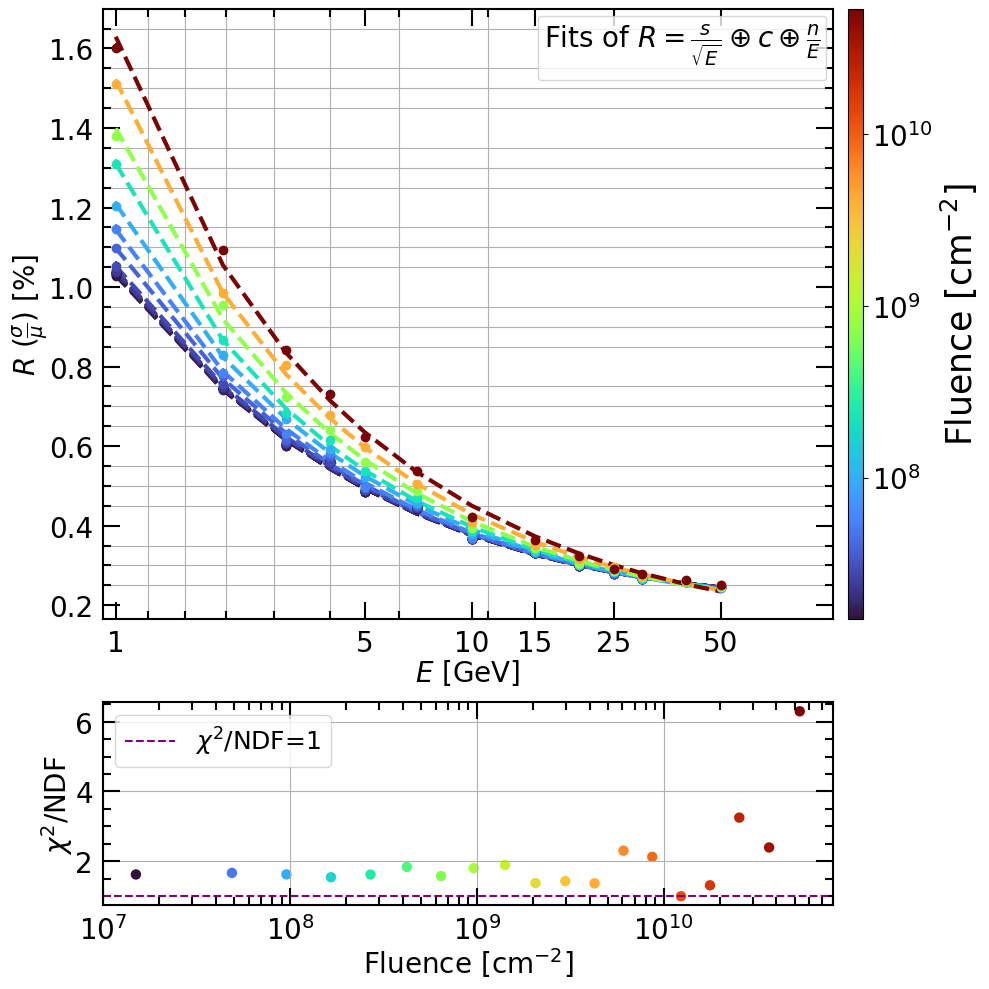}}
\hfill

\caption{Calorimeter linearity (left) and resolution (right) versus DCR, expressed as fluence per Fig.~\ref{fig:fig_dcrtofluence}. The left plot's main graph shows the Crystal Ball mean, $\mu$ versus particle energy, with the top inset showing percentage deviation from linearity. The dashed purple line represents linearity, $\mu=E$. The right plot's main graph depicts resolution, $\frac{\sigma}{\mu}$, versus energy in inverse square-root scaling, with dashed lines indicating fits of Eq.~\ref{eq:eq_resolution}. The bottom inset shows reduced $\chi^{2}$ of the fits versus fluence. The colour scale for both figures indicates fluence from low (blue) to high (red). Values with $DCR$ outside the valid range in Fig.~\ref{fig:fig_dcrtofluence} cannot be shown.}

\label{fig:fig_linres}
\end{figure}

\begin{figure}[htpb]
    \centering
    \includegraphics[width=0.49\linewidth]{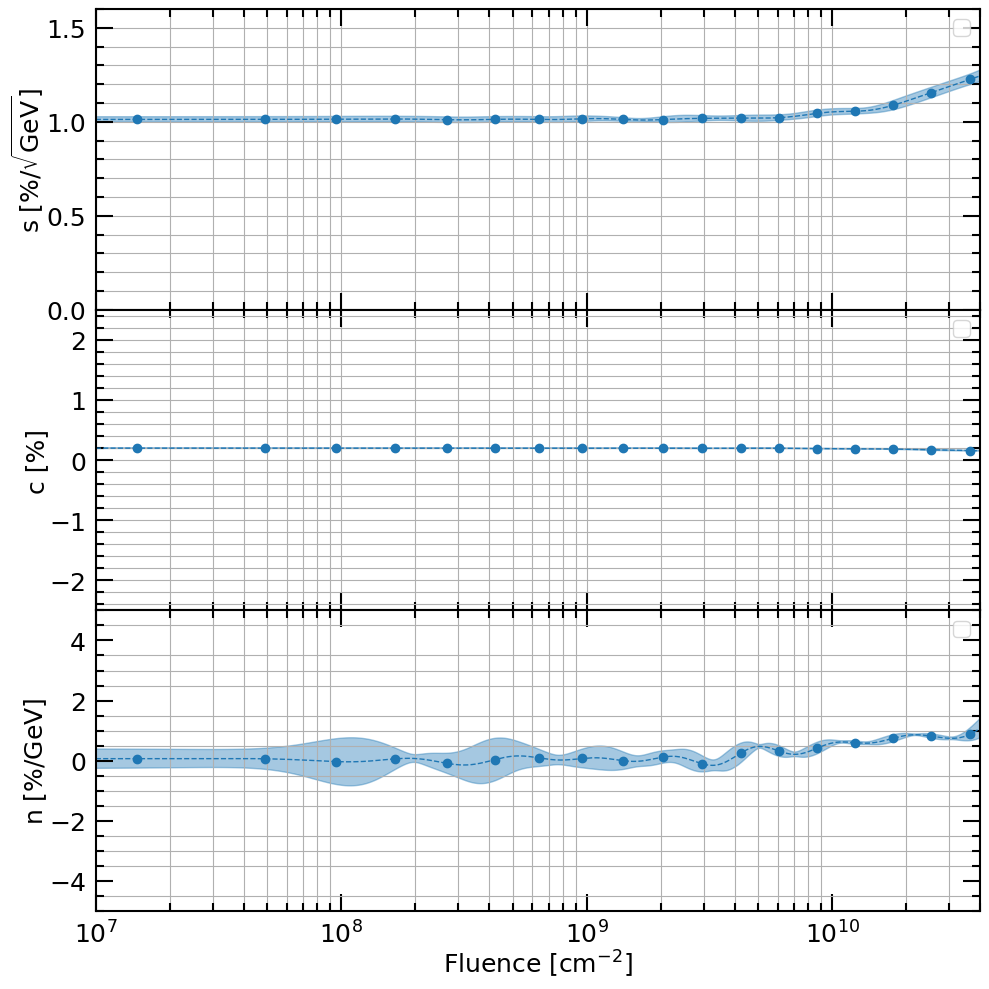}
    \caption{Fitted parameters of Eq.~\ref{eq:eq_resolution} as a function of fluence. The blue points indicate the values obtained from a fit. The dashed lines and shaded regions indicate the spline fit used to obtain the fluence and its 1-$\sigma$ confidence band. Values with $DCR$ outside the valid range in Fig.~\ref{fig:fig_dcrtofluence} cannot be shown. }
    \label{fig:fig_resolution_resparams}
\end{figure}

Fig.~\ref{fig:fig_linearity} shows that as fluence increases, there is an increasing systematic bias to the energy reconstructed by the calorimeter. This is indicated by the severe deterioration of the linearity at energies in the range \SIrange{1}{10}{\giga \electronvolt}. For fluences greater than $\qty{1e10}{\per \centi \meter \squared}$, the deviation from linearity is severe. At the $DCR=\qty{1.585}{\giga \hertz}$, a \qty{1}{\giga \electronvolt} shower is overestimated in energy by \qty{45}{\percent}, which is also shown in Fig.~\ref{fig:fig_energysums_cut_1gev}. The reasons for this effect are discussed in previous sections. The intrinsic \qtyrange{6}{8}{\percent} deviation from linearity at the lowest $DCR$ value, shown in deep indigo, is due to the cross-talk included in the digitisation, as previously mentioned. 

Fig.~\ref{fig:fig_resolution} shows the resolution of the calorimeter as a function of fluence. The bottom inset indicates that for fluences below around $\qty{5e10}{\per \centi \meter \squared}$, the fit quality is generally acceptable ($\frac{\chi^{2}}{\mathrm{NDF}} \lesssim 2$). However, the last three data points show poorer agreement with the resolution model, ($2 < \frac{\chi^{2}}{\mathrm{NDF}} \leq 5$. This can be attributed to the bias introduced by the presence of the cut. Despite this unavoidable effect, there is nonetheless qualitatively acceptable agreement with the resolution model.  The main figure demonstrates that the resolution does not experience a significant degradation until fluences greater than \qty{1e10}{\per \centi \meter \squared}

Fig.~\ref{fig:fig_resolution_resparams} illustrates that the degradation in performance is expressed in terms of a combination of the stochastic resolution and noise term. The stochastic resolution of the calorimeter remains stable up until around \qty{1e10}{\per \centi \meter \squared} fluence and increases by no more than \qty{0.2}{\percent \per \sqrt \giga \electronvolt} over the range of studied dark count rates after around \qty{1e10}{\per \centi \meter \squared} fluence. The noise term also increases marginally over the same range by around \qty{1}{\percent \per \giga \electronvolt}. However, the value is consistent with the fitted value at fluences lower than \qty{1e10}{\per \centi \meter \squared} due to a large error. As expected, the constant term, $c$, is not affected by the presence of detector noise.

In summary, the presented results indicate the resolution ought to remain unaffected by the irradiation up to fluences of \qty{1e10}{\per \centi \meter \squared}. The main challenge posed by dark noise is instead the systematic biases introduced to the mean energy due to the presence of the noise cut.

\subsection{Temperature Stability}

\begin{figure}[htpb]
    \centering
    \includegraphics[width=0.49\linewidth]{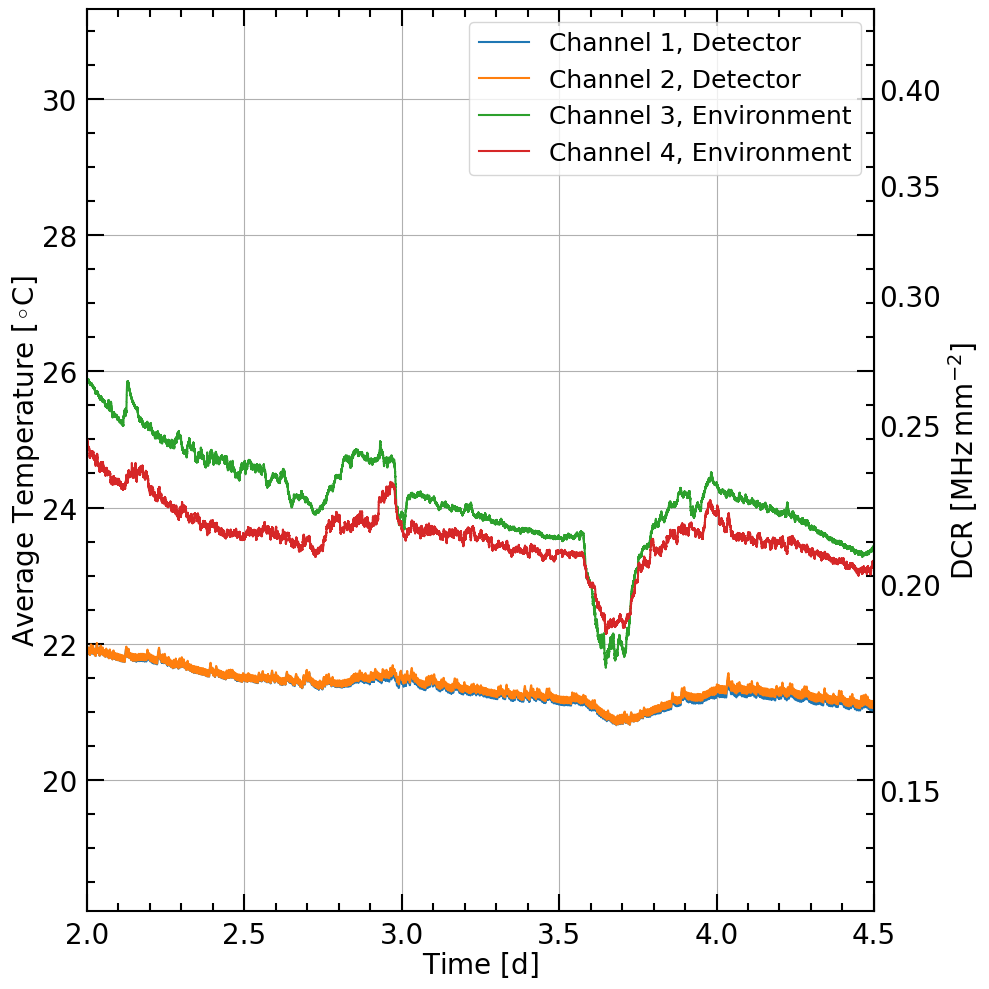}
    \caption{Average temperature of a module and the environment during a 2024 testbeam at the Super Proton Synchrotron at CERN, Geneva, Switzerland. The blue and orange lines indicate the temperature of the module measured by two channels, while the red and green lines indicate the ambient temperature. The $x$-axis shows the time of the measurement in days. The left $y$-axis shows the average temperature, and the right $y$-axis shows the $DCR$ expected at that temperature at zero fluence from the relation shown in Fig.~\ref{fig:fig_temperaturetodcrratio} and with $DCR(\Phi = 0, T = \qty{20}{\celsius}) = \qty{0.152(0.111)}{\mega \hertz \per \milli \meter \squared}$. The error cannot be shown on the axis.}
    \label{fig:fig_avgtemptodcr}
\end{figure}

\begin{figure}[htpb]
\subfloat[\label{fig:fig_stemperaturefluctuations}]{\includegraphics[width=0.49\linewidth]{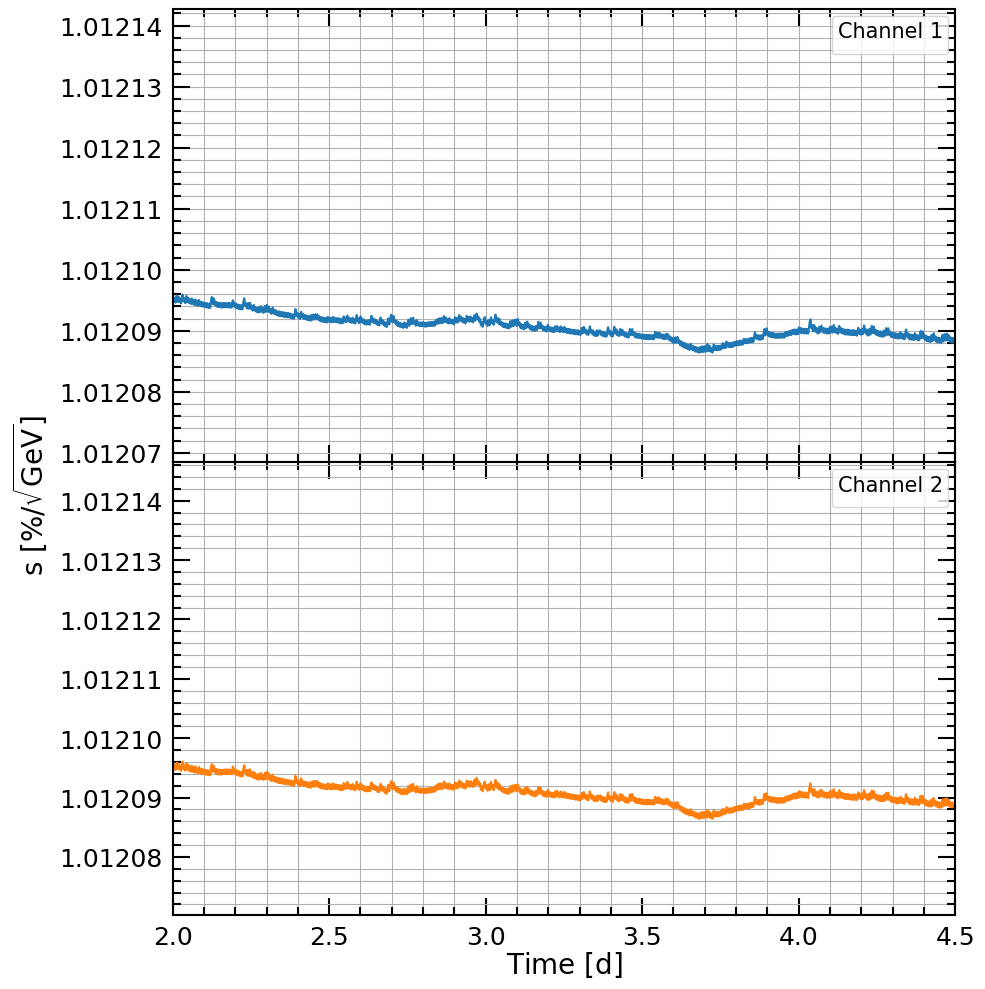}}
\subfloat[ \label{fig:fig_ntemperaturefluctuations}]   {\includegraphics[width=0.49\linewidth]{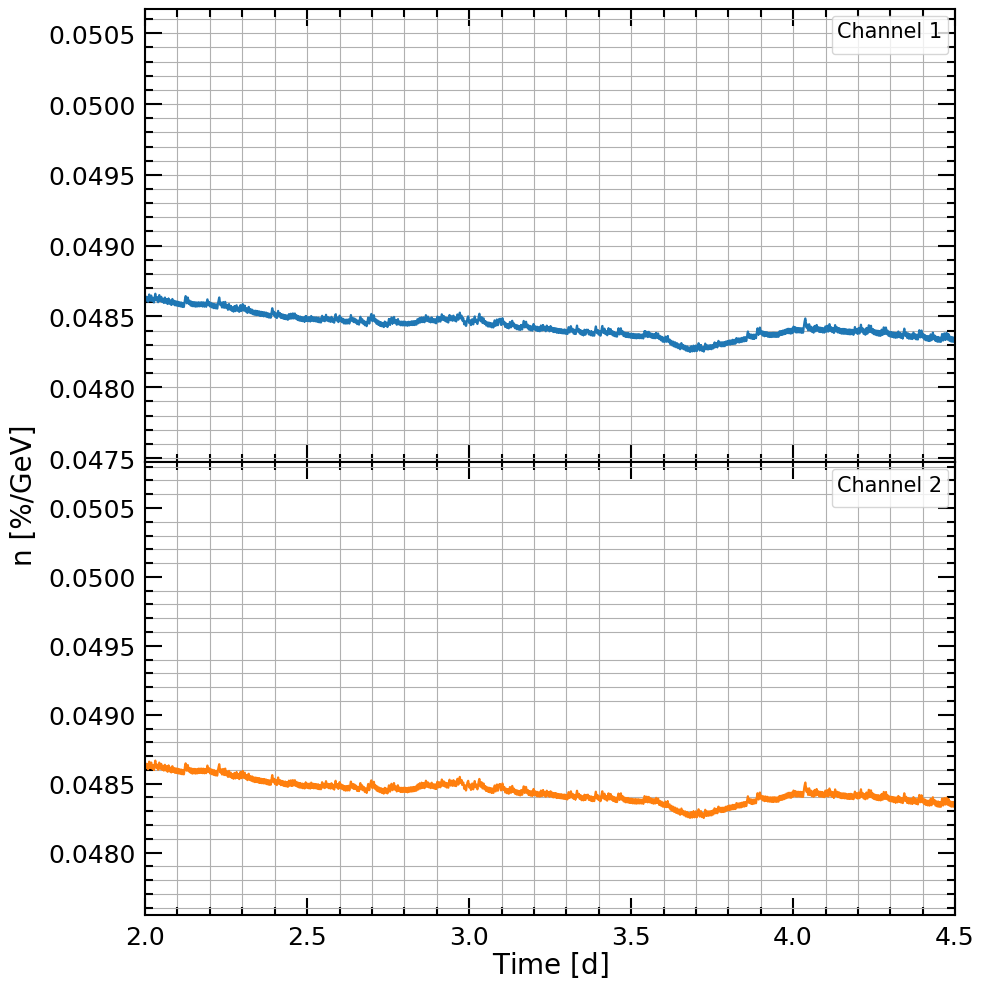}}
\hfill

\caption{Fluctuations in the stochastic and noise resolution parameters obtained from the fits shown in Fig.~\ref{fig:fig_resolution} and Fig.~\ref{fig:fig_resolution_resparams}, using the temperature relation shown in Fig.~\ref{fig:fig_temperaturetodcrratio} and with $DCR(\Phi = 0, T = \qty{20}{\celsius}) = \qty{0.152(0.111)}{\mega \hertz \per \milli \meter \squared}$. The stochastic and noise terms are shown in the left and right plots. The blue and orange lines show the results for Channel 1 and Channel 2 respectively. Error bands are too large to be shown.}

\label{fig:fig_temperaturefluctuations}
\end{figure}

Fig.~\ref{fig:fig_avgtemptodcr} shows the average testbeam temperatures for the testbeam period from the start of day 2 to the middle of day 4 of data-taking, with the corresponding DCR obtained using the green dashed line of Fig.~\ref{fig:fig_temperaturetodcrratio} for the module and environment. This range was selected as it was the most stable period of data-taking in terms of temperature. The average temperature is close to \qty{20}{\celsius}, which is similar to the ranges studied in the available literature measurements referenced in Sec.~\ref{sec:DCR_vs_fluence} and Sec.~\ref{sec:DCR_vs_T}. The blue and orange lines indicate the corresponding $DCR$ is expected to vary by no more than around \qty{50}{\kilo \hertz \per \milli \meter \squared} for a \qtyproduct{3x3}{\milli \meter \squared} sensor. For non-irradiated sensors, examination of Fig.~\ref{fig:fig_resolution_resparams} at the minimum studied fluence at $\Phi = \qty{1.39e6}{\per \centi \meter \squared}$ indicates negligible influence of temperature on the calorimeter's resolution or linearity.

 Using an interpolating spline of the fitted values of the stochastic and constant terms shown in Fig.~\ref{fig:fig_resolution_resparams} as a function of $DCR$, the expected fluctuations in these parameters can be determined for the module, and are shown in Fig.~\ref{fig:fig_temperaturefluctuations}. It can be concluded that the fluctuations are many orders of magnitude smaller than the value. Therefore, the resolution is expected to remain essentially unaffected by temperature fluctuations given the assumptions made in this study.  As a caveat, it is unknown if this will remain the case after irradiation.

\section{Conclusion}

In this study, a simulation of electron showers and SiPM dark noise were combined to estimate the degradation of performance expected from the SiPM readout of a highly granular crystal calorimeter designed for Particle Flow. The effect on the energy response of single sensors, the module, the false positive rate of the trigger and the linearity and resolution of the calorimeter were studied as a function of increasing dark noise, related to fluence using the results of a study presented in Ref.~\cite{altamura_radiation_2023}. The influence of the temperature on the resolution is also studied using the results, using the temperature fluctuations obtained from a 2024 testbeam.

The study indicates that the primary degradation of performance that will be experienced by the calorimeter manifests as degrading resolution and a systematic bias in energy due to the biasing of the hit energy distribution caused by noise suppression cuts. These effects will degrade the calorimeter performance most significantly for showers with energy in the range \SIrange{1}{10}{\giga \electronvolt}. This results in a severe deviation from linearity in this energy range. However, the energy independence of the scale of the noise means that the overall resolution of the calorimeter is not strongly affected by the presence of dark noise, with stochastic resolution increasing by only \qty{0.5}{\percent \per \sqrt \giga \electronvolt} due to dark noise. The effectiveness of the noise cut applied also decreases with fluence.  The calorimeter resolution is demonstrated to remain stable and generally unaffected by dark noise up to around a fluence of \qty{1e10}{\per \centi \meter \squared}, after which performance begins to degrade. In practice, this limit will likely be higher due to the assumption made in this study that the entire calorimeter is uniformly irradiated. Additionally, under the assumption of non-irradiation, it is found that the calorimeter resolution will be essentially unaffected by temperature fluctuations during operation.

The study should be interpreted with caveats. It assumes uniform irradiation of the detector when radiation damage will vary with proximity to the beam and the pseudorapidity range covered by instrumentation. Irradiation also causes correlated noise from SiPM discharges, which is challenging to measure accurately for irradiated devices \cite{garutti_afterpulse_2014}. Relationships between dark count rate, fluence, and temperature are based on existing studies, as the specific SiPM for the calorimeter is not yet chosen. Finally, the relationship of radiation damage with fluence and temperature is complex, and therefore each effect is studied while holding the other constant. Results should be interpreted with these limitations in mind.

Overall, this study illustrates the robustness of the calorimeter system to irradiation and temperature. Nonetheless, it also highlights challenges that must be overcome to achieve excellent electromagnetic resolution in a Particle Flow detector system using a crystal calorimeter.

\appendix

\section{Simulation Parameters}

\begin{table}[h]
    \centering
     \caption{Table of SiPM simulation parameters for the program of Ref.~ \cite{garutti_simulation_2021} and their meanings, units, and values. In this table, $G$ is not an input into the simulation program, and is presented for reference. $\varepsilon$ indicates a negligbly small number that is used instead of zero due to the limitations of the simulation program.}
    \label{tab:SiPMSimulationParameters}
    \begin{small}
    \begin{tabular}{c l c c}
        \toprule
        \textbf{Parameter} & \textbf{Meaning} & \textbf{Unit} & \textbf{Value} \\ \midrule
        Area & Surface Area of SiPM & \unit{\milli \meter \squared} &  \numproduct{3 x 3} \\
        $G$  & Gain/Amplification Factor & n.p.e & \qty{1e5}{}  \\
        $t_0$ & Integration Time before Start of Gate & ns & 100 \\ 
        $\sigma_0$ & Pedestal Width & G\textsuperscript{-1} & 0.075 \\ 
        $\sigma_1$ & Gain Smearing & G\textsuperscript{-1} & 0.02 \\
        $t_{\text{start}}$ & Start of Gate & \unit{\nano \second} & -5 \\ 
        $t_{\text{gate}}$ & Gate Length & \unit{\nano \second} & 100 \\ 
        $p_{\text{pXT}}$ & Prompt Cross Talk Probability &  & 0.12 \\
        $p_{\text{dXT}}$ & Delayed Cross Talk Probability &  & 0 \\ 
        $\mu$ & Mean \# of Geiger Discharges from $\gamma$ &  & 
        $\varepsilon$ \\
        $\tau_{\text{rec}}$ & Recovery Time Of SiPM & \unit{\nano \second} & 20 \\
        $\tau_{\text{dXT}}$ & Delayed Cross Talk Recovery Time & \unit{\nano \second} & 25 \\
        $p_{\text{Ap}}$ & Afterpulse Probability &  & 0.01 \\ 
        $\tau_{\text{Ap}}$ & Afterpulse Time Constant & \unit{\nano \second} & 7.5 \\ 
        $\tau_{\text{s}}$ & Slow Component of SiPM Pulse & \unit{\nano \second} & 20.0 \\
        $r_{\text{f}}$ & Fraction of Fast Pulse &  & $\varepsilon$ \\
        $DCR$ & Dark Count Rate & GHz & \qty{1}{\mega \hertz} - \qty{1.58}{\giga \hertz}, log scale \\
        \bottomrule
    \end{tabular}
    \end{small}
   
\end{table}

\section{Relationship Between Dark Count Rate and Fluence}\label{sec:DCR_vs_fluence}

In Ref.~\cite{altamura_radiation_2023}, a study was conducted to irradiate SiPMs manufactured by HPK with protons of energy \qty{47}{\mega \electronvolt} protons in a range of neutron-equivalent fluences from \qtyrange{7.4e6}{6.4e11}{\per \centi \meter \squared}. The dark count rate ($DCR$) was measured at 12 stages of irradiation within this range. However, uncertainties were not reported in the study. DCR was extracted from the dark current of the SiPM under test and 'current gain', which corresponds to the ratio between the average charges induced by a single discharge with all correlated noise effects to the average charge without them. This method was chosen due to the significant noise levels of the SiPM after irradiation. The measurements of DCR were performed at \qty{20}{ \celsius}. \cite{acerbi_private}. Further details are available in the reference.

The results of this study were used to estimate the relationship between the SiPM dark count rate and fluence. Since the specific SiPM that will be used for the crystal calorimeter readout is not yet ascertained, an attempt was made to capture the overall trend of the measured data.

\begin{figure}

\subfloat[\label{fig:fig_fluencetodcr}]{\includegraphics[width=0.49\linewidth]{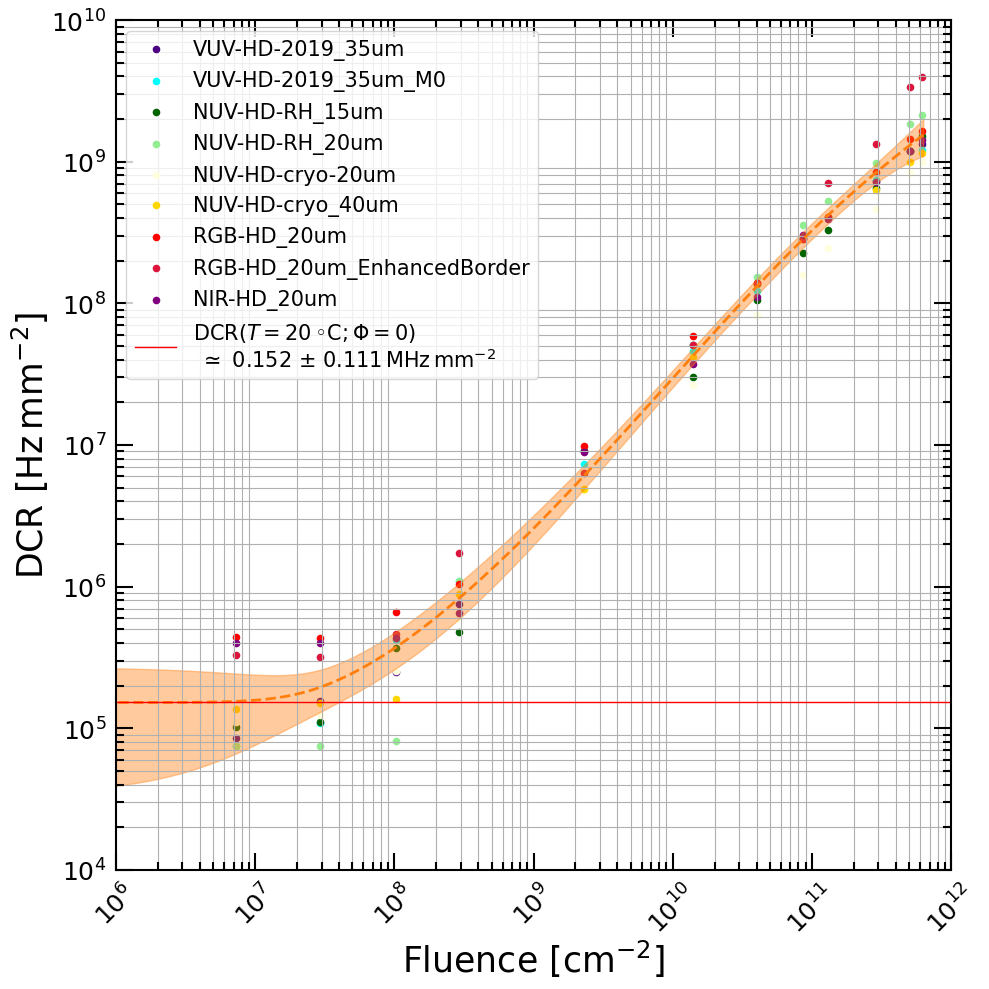}}
\hfill
\subfloat[ \label{fig:fig_dcrtofluence}] {\includegraphics[width=0.49\linewidth]{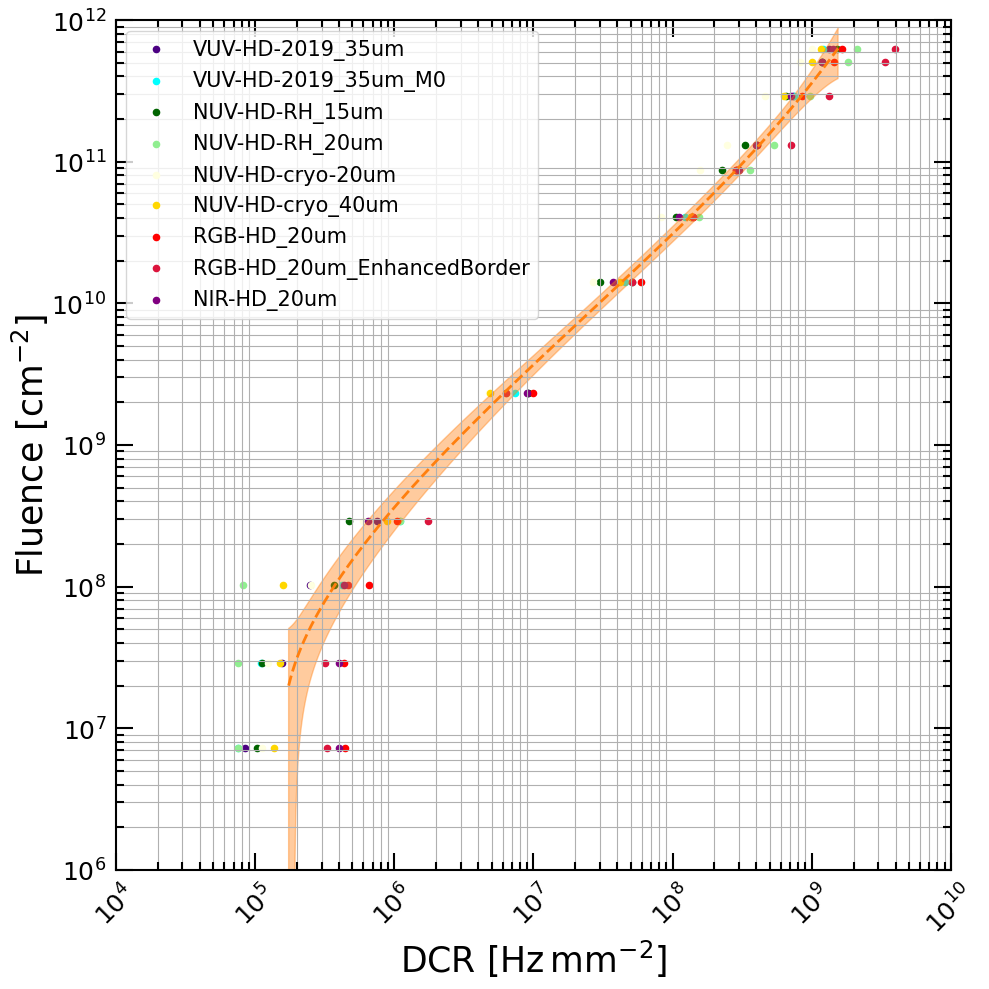}}

\caption{Extracted relationships between DCR and fluence from Ref.~\cite{altamura_radiation_2023}. Data was extracted from Fig.~12 of the reference. Fig.~\ref{fig:fig_fluencetodcr} shows DCR versus fluence as presented in the reference and Fig.~\ref{fig:fig_dcrtofluence} shows fluence versus DCR, obtained by inversion. The coloured dots and legend represent the various SiPMs under test, with details of their properties available in the reference. The orange dashed line and the shaded orange shaded area indicate the bootstrap mean and the 1-$\sigma$ uncertainty bands, respectively. In Fig.~\ref{fig:fig_fluencetodcr}, the turning point is indicated by the red line, and is used to estimate the non-irradiated DCR.  }

\label{fig:fig_fluencedcrsplines}

\end{figure}

Firstly, for \num{1e5} iterations, a data point was randomly selected from each fluence bin, and a smoothed spline fit was made between the points. Then, the bootstrapped mean and standard deviation of the \num{1e5} spline fits were used to estimate the value and uncertainty of the DCR as a function of fluence, shown in Fig.~\ref{fig:fig_fluencetodcr}. The bootstrap spline was then inverted to  estimate fluence as a function of DCR. The estimate and the uncertainty are shown in Fig.~\ref{fig:fig_dcrtofluence}.

As a caveat, it should be assumed that the conversion from dark count rate to fluence in this study is approximate. Additionally, Fig.~\ref{fig:fig_dcrtofluence} shows that the dark count rate (DCR) becomes independent of the fluence at levels below \qty{1e7}{\per \centi \meter \squared}. The DCR of the non-irradiated devices can be inferred from the fluences below this limit \cite{acerbi_private}. Consequently, DCR values below around \qty{100}{\kilo \hertz \per \milli \meter \squared} cannot be used to estimate fluence rates.

The turning point of the spline at \qty{1.39e6}{\per \centi \meter \squared} fluence was used as an estimate for the non-irradiated value of DCR. It was estimated to be $DCR(\Phi = 0, T = \qty{20}{\celsius}) = \qty{0.152(0.111)}{\mega \hertz \per \milli \meter \squared}$ using this method.

\section{Relationship between Dark Count Rate and Temperature}\label{sec:DCR_vs_T}

A study presented in Ref.~\cite{otte_characterization_2017} describes the relationship between the DCR of three types of non-irradiated SiPM as a function of temperature. This measurement was achieved by the analysis of SiPM signals of devices measured without illumination in a climate chamber at temperatures of \qty{-40}{\celsius}, \qty{-20}{\celsius}, \qty{0}{\celsius}, \qty{20}{\celsius} and \qty{40}{\celsius}. Explicitly, the study counts the ratio of the number of signals from the SiPM with an amplitude surpassing \qty{0.5}{\mathrm{p.e}} to the duration of all the voltage pulses in the dark. Pulses were measured at \qty{90}{\percent} breakdown probability, with a minor contribution of correlated noise to the $DCR$ of \qty{2}{\percent}. Such a measurement is possible compared to the study of Ref.~\cite{altamura_radiation_2023} due to the significantly lower noise rate ($DCR < \qty{1}{\kilo \hertz \per \milli \meter \squared}$). Further details of the study are available in the reference. 

\begin{figure}[htpb]
    \centering
    \includegraphics[width=0.49\linewidth]{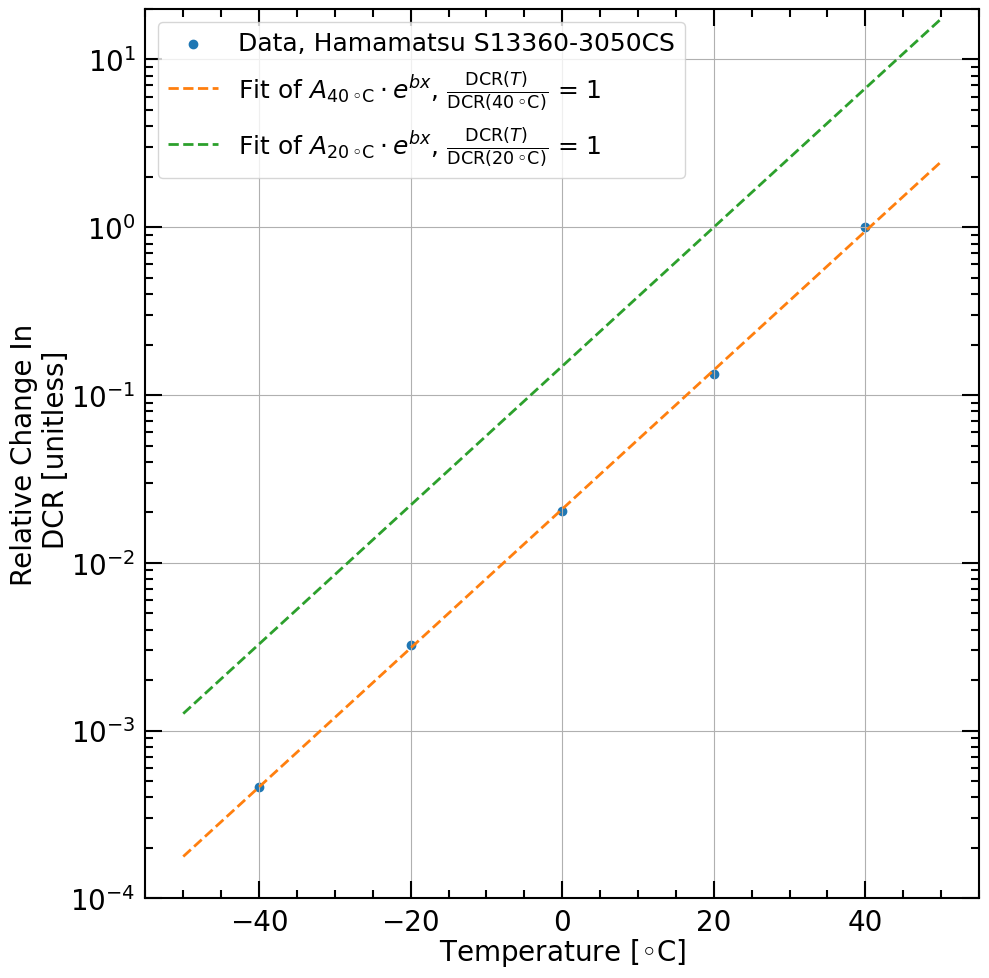}
    \caption{The ratio of DCR to DCR at \qty{40}{\celsius} (\qty{20}{\celsius}) as a function of temperature. The blue points represent the data from Fig. 23b of Ref.~\cite{otte_characterization_2017}. The orange dashed line indicates a fit of Eq.~\ref{eq:DCRT} to the data. The green dashed line indicates the same fit as the orange dashed line, re-scaled so that the ratio is 1 at $T = \qty{20}{\celsius}$.}
    \label{fig:fig_temperaturetodcrratio}
\end{figure}

The ratio of the DCR to the DCR measured at \qty{40}{\celsius} from that study is shown in Fig. for a Hamamatsu S13360-3050CS. This SiPM was chosen because it has a \qtyproduct{3x3}{\milli \meter \squared} area, which is the same as expected to be used in the crystal calorimeter readout. Ref.~\cite{otte_characterization_2017} observes that the relationship is well described by the Eq.~\ref{eq:DCRT}:

\begin{equation}
 \frac{DCR(\Phi=0, T)}{DCR(\Phi=0, T=\qty{40}{\celsius})} = A e^{bT} 
 \label{eq:DCRT}
\end{equation}

As in the reference, a fit using least squares regression was made of the extracted data using Eq.~\ref{eq:DCRT}. The agreement was observed to be within around \qty{6}{\percent} of the measured value. However, visible errors were not presented in the reference and therefore a $\chi^{2}$ goodness-of-fit value cannot be shown. The proportional change of DCR can then be re-scaled to be relative to the DCR at $T=\qty{20}{\celsius}$. This can then be multiplied by the value obtained for $DCR(\Phi = 0, T = \qty{20}{\celsius})$ in Sec.~\ref{sec:DCR_vs_fluence} to obtain an estimate for DCR as a function of temperature at \qty{20}{\celsius}.

\newpage

\nocite{*}

\end{document}